\documentclass[longauth]{aa}
\usepackage[utf8]{inputenc}
\usepackage[varg]{txfonts}
\usepackage{hyperref}
\usepackage{graphicx}
\usepackage{natbib}
\usepackage{amssymb}

\usepackage{scalerel}
\usepackage{euclid}

\usepackage{xcolor}

\usepackage{hyperref}
\hypersetup{
	colorlinks=true,
	linkcolor=blue,
	filecolor=magenta,      
	urlcolor=blue,
	citecolor=blue
}
\makeatletter
\renewcommand*\aa@pageof{, page \thepage{} of \pageref*{LastPage}}
\makeatother

\newcommand{\rom}[1]{\uppercase\expandafter{\romannumeral #1\relax}}

\bibpunct{(}{)}{;}{a}{}{,} 

\newcommand{\phz}{photometric redshift}
\newcommand{\phdz}{photometric-redshift}
\newcommand{\dd}{\mathrm{d}}

\begin{document} 

\title{\Euclid preparation. XXXI. The effect of the variations in photometric passbands on photometric-redshift accuracy}

\newcommand{\orcid}[1]{} 
\author{Euclid Collaboration: S.~Paltani\orcid{0000-0002-8108-9179}$^{1}$\thanks{\email{stephane.paltani@unige.ch}}, J.~Coupon$^{1}$, W.~G.~Hartley$^{1}$, A.~Alvarez-Ayllon$^{1}$, F.~Dubath\orcid{0000-0002-6533-2810}$^{1}$, J.J.~Mohr\orcid{0000-0002-6875-2087}$^{2,3}$, M.~Schirmer\orcid{0000-0003-2568-9994}$^{4}$, J.-C.~Cuillandre\orcid{0000-0002-3263-8645}$^{5}$, G.~Desprez$^{1,6}$, O.~Ilbert\orcid{0000-0002-7303-4397}$^{7}$
% ...
}

\institute{$^{1}$ Department of Astronomy, University of Geneva, ch. d'\'Ecogia 16, 1290 Versoix, Switzerland\\
	$^{2}$ University Observatory, Faculty of Physics, Ludwig-Maximilians-Universit{\"a}t, Scheinerstr. 1, 81679 Munich, Germany\\
	$^{3}$ Max Planck Institute for Extraterrestrial Physics, Giessenbachstr. 1, 85748 Garching, Germany\\
	$^{4}$ Max-Planck-Institut f\"ur Astronomie, K\"onigstuhl 17, 69117 Heidelberg, Germany\\
	$^{5}$ Universit\'e Paris-Saclay, Universit\'e Paris Cit\'e, CEA, CNRS, Astrophysique, Instrumentation et Mod\'elisation Paris-Saclay, 91191 Gif-sur-Yvette, France\\
	$^{6}$ Department of Astronomy \& Physics and Institute for Computational Astrophysics, Saint Mary's University, 923 Robie Street, Halifax, Nova Scotia, B3H 3C3, Canada\\
	$^{7}$ Aix-Marseille Universit\'e, CNRS, CNES, LAM, Marseille, France\\
	$^{8}$ Leiden Observatory, Leiden University, Niels Bohrweg 2, 2333 CA Leiden, The Netherlands\\
	$^{9}$ Universit\'e Paris-Saclay, CNRS, Institut d'astrophysique spatiale, 91405, Orsay, France\\
	$^{10}$ ESAC/ESA, Camino Bajo del Castillo, s/n., Urb. Villafranca del Castillo, 28692 Villanueva de la Ca\~nada, Madrid, Spain\\
	$^{11}$ Institute of Cosmology and Gravitation, University of Portsmouth, Portsmouth PO1 3FX, UK\\
	$^{12}$ INAF-Osservatorio di Astrofisica e Scienza dello Spazio di Bologna, Via Piero Gobetti 93/3, 40129 Bologna, Italy\\
	$^{13}$ Dipartimento di Fisica e Astronomia "Augusto Righi" - Alma Mater Studiorum Universit\'a di Bologna, via Piero Gobetti 93/2, 40129 Bologna, Italy\\
	$^{14}$ INFN-Sezione di Bologna, Viale Berti Pichat 6/2, 40127 Bologna, Italy\\
	$^{15}$ Universit\"ats-Sternwarte M\"unchen, Fakult\"at f\"ur Physik, Ludwig-Maximilians-Universit\"at M\"unchen, Scheinerstrasse 1, 81679 M\"unchen, Germany\\
	$^{16}$ INAF-Osservatorio Astrofisico di Torino, Via Osservatorio 20, 10025 Pino Torinese (TO), Italy\\
	$^{17}$ Dipartimento di Fisica, Universit\'a di Genova, Via Dodecaneso 33, 16146, Genova, Italy\\
	$^{18}$ INFN-Sezione di Roma Tre, Via della Vasca Navale 84, 00146, Roma, Italy\\
	$^{19}$ Department of Physics "E. Pancini", University Federico II, Via Cinthia 6, 80126, Napoli, Italy\\
	$^{20}$ INAF-Osservatorio Astronomico di Capodimonte, Via Moiariello 16, 80131 Napoli, Italy\\
	$^{21}$ Instituto de Astrof\'isica e Ci\^encias do Espa\c{c}o, Universidade do Porto, CAUP, Rua das Estrelas, PT4150-762 Porto, Portugal\\
	$^{22}$ Dipartimento di Fisica, Universit\'a degli Studi di Torino, Via P. Giuria 1, 10125 Torino, Italy\\
	$^{23}$ INFN-Sezione di Torino, Via P. Giuria 1, 10125 Torino, Italy\\
	$^{24}$ INAF-IASF Milano, Via Alfonso Corti 12, 20133 Milano, Italy\\
	$^{25}$ INAF-Osservatorio Astronomico di Roma, Via Frascati 33, 00078 Monteporzio Catone, Italy\\
	$^{26}$ INFN-Sezione di Roma, Piazzale Aldo Moro, 2 - c/o Dipartimento di Fisica, Edificio G. Marconi, 00185 Roma, Italy\\
	$^{27}$ Institut de F\'{i}sica d'Altes Energies (IFAE), The Barcelona Institute of Science and Technology, Campus UAB, 08193 Bellaterra (Barcelona), Spain\\
	$^{28}$ Port d'Informaci\'{o} Cient\'{i}fica, Campus UAB, C. Albareda s/n, 08193 Bellaterra (Barcelona), Spain\\
	$^{29}$ Institut d'Estudis Espacials de Catalunya (IEEC), Carrer Gran Capit\'a 2-4, 08034 Barcelona, Spain\\
	$^{30}$ Institute of Space Sciences (ICE, CSIC), Campus UAB, Carrer de Can Magrans, s/n, 08193 Barcelona, Spain\\
	$^{31}$ INFN section of Naples, Via Cinthia 6, 80126, Napoli, Italy\\
	$^{32}$ Centre National d'Etudes Spatiales -- Centre spatial de Toulouse, 18 avenue Edouard Belin, 31401 Toulouse Cedex 9, France\\
	$^{33}$ Institut national de physique nucl\'eaire et de physique des particules, 3 rue Michel-Ange, 75794 Paris C\'edex 16, France\\
	$^{34}$ Institute for Astronomy, University of Edinburgh, Royal Observatory, Blackford Hill, Edinburgh EH9 3HJ, UK\\
	$^{35}$ Jodrell Bank Centre for Astrophysics, Department of Physics and Astronomy, University of Manchester, Oxford Road, Manchester M13 9PL, UK\\
	$^{36}$ European Space Agency/ESRIN, Largo Galileo Galilei 1, 00044 Frascati, Roma, Italy\\
	$^{37}$ University of Lyon, Univ Claude Bernard Lyon 1, CNRS/IN2P3, IP2I Lyon, UMR 5822, 69622 Villeurbanne, France\\
	$^{38}$ Institute of Physics, Laboratory of Astrophysics, Ecole Polytechnique F\'ed\'erale de Lausanne (EPFL), Observatoire de Sauverny, 1290 Versoix, Switzerland\\
	$^{39}$ Mullard Space Science Laboratory, University College London, Holmbury St Mary, Dorking, Surrey RH5 6NT, UK\\
	$^{40}$ Departamento de F\'isica, Faculdade de Ci\^encias, Universidade de Lisboa, Edif\'icio C8, Campo Grande, PT1749-016 Lisboa, Portugal\\
	$^{41}$ Instituto de Astrof\'isica e Ci\^encias do Espa\c{c}o, Faculdade de Ci\^encias, Universidade de Lisboa, Campo Grande, 1749-016 Lisboa, Portugal\\
	$^{42}$ INFN-Padova, Via Marzolo 8, 35131 Padova, Italy\\
	$^{43}$ INAF-Osservatorio Astronomico di Trieste, Via G. B. Tiepolo 11, 34143 Trieste, Italy\\
	$^{44}$ Aix-Marseille Universit\'e, CNRS/IN2P3, CPPM, Marseille, France\\
	$^{45}$ Istituto Nazionale di Fisica Nucleare, Sezione di Bologna, Via Irnerio 46, 40126 Bologna, Italy\\
	$^{46}$ INAF-Osservatorio Astronomico di Padova, Via dell'Osservatorio 5, 35122 Padova, Italy\\
	$^{47}$ Institute of Theoretical Astrophysics, University of Oslo, P.O. Box 1029 Blindern, 0315 Oslo, Norway\\
	$^{48}$ Technical University of Denmark, Elektrovej 327, 2800 Kgs. Lyngby, Denmark\\
	$^{49}$ Cosmic Dawn Center (DAWN), Denmark\\
	$^{50}$ Institut d'Astrophysique de Paris, 98bis Boulevard Arago, 75014, Paris, France\\
	$^{51}$ Jet Propulsion Laboratory, California Institute of Technology, 4800 Oak Grove Drive, Pasadena, CA, 91109, USA\\
	$^{52}$ Universit\'e de Gen\`eve, D\'epartement de Physique Th\'eorique and Centre for Astroparticle Physics, 24 quai Ernest-Ansermet, CH-1211 Gen\`eve 4, Switzerland\\
	$^{53}$ Department of Physics, P.O. Box 64, 00014 University of Helsinki, Finland\\
	$^{54}$ Helsinki Institute of Physics, Gustaf H{\"a}llstr{\"o}min katu 2, University of Helsinki, Helsinki, Finland\\
	$^{55}$ NOVA optical infrared instrumentation group at ASTRON, Oude Hoogeveensedijk 4, 7991PD, Dwingeloo, The Netherlands\\
	$^{56}$ Argelander-Institut f\"ur Astronomie, Universit\"at Bonn, Auf dem H\"ugel 71, 53121 Bonn, Germany\\
	$^{57}$ Department of Physics, Institute for Computational Cosmology, Durham University, South Road, DH1 3LE, UK\\
	$^{58}$ Infrared Processing and Analysis Center, California Institute of Technology, Pasadena, CA 91125, USA\\
	$^{59}$ Universit\'e C\^{o}te d'Azur, Observatoire de la C\^{o}te d'Azur, CNRS, Laboratoire Lagrange, Bd de l'Observatoire, CS 34229, 06304 Nice cedex 4, France\\
	$^{60}$ Institut d'Astrophysique de Paris, UMR 7095, CNRS, and Sorbonne Universit\'e, 98 bis boulevard Arago, 75014 Paris, France\\
	$^{61}$ Universit\'e Paris Cit\'e, CNRS, Astroparticule et Cosmologie, 75013 Paris, France\\
	$^{62}$ University of Applied Sciences and Arts of Northwestern Switzerland, School of Engineering, 5210 Windisch, Switzerland\\
	$^{63}$ European Space Agency/ESTEC, Keplerlaan 1, 2201 AZ Noordwijk, The Netherlands\\
	$^{64}$ Department of Physics and Astronomy, University of Aarhus, Ny Munkegade 120, DK-8000 Aarhus C, Denmark\\
	$^{65}$ Centre for Astrophysics, University of Waterloo, Waterloo, Ontario N2L 3G1, Canada\\
	$^{66}$ Department of Physics and Astronomy, University of Waterloo, Waterloo, Ontario N2L 3G1, Canada\\
	$^{67}$ Perimeter Institute for Theoretical Physics, Waterloo, Ontario N2L 2Y5, Canada\\
	$^{68}$ Space Science Data Center, Italian Space Agency, via del Politecnico snc, 00133 Roma, Italy\\
	$^{69}$ Institute of Space Science, Str. Atomistilor, nr. 409 M\u{a}gurele, Ilfov, 077125, Romania\\
	$^{70}$ Instituto de Astrof\'isica de Canarias, Calle V\'ia L\'actea s/n, 38204, San Crist\'obal de La Laguna, Tenerife, Spain\\
	$^{71}$ Departamento de Astrof\'isica, Universidad de La Laguna, 38206, La Laguna, Tenerife, Spain\\
	$^{72}$ Dipartimento di Fisica e Astronomia "G. Galilei", Universit\'a di Padova, Via Marzolo 8, 35131 Padova, Italy\\
	$^{73}$ Dipartimento di Fisica e Astronomia, Universit\'a di Bologna, Via Gobetti 93/2, 40129 Bologna, Italy\\
	$^{74}$ Departamento de F\'isica, FCFM, Universidad de Chile, Blanco Encalada 2008, Santiago, Chile\\
	$^{75}$ Centre for Electronic Imaging, Open University, Walton Hall, Milton Keynes, MK7~6AA, UK\\
	$^{76}$ AIM, CEA, CNRS, Universit\'{e} Paris-Saclay, Universit\'{e} de Paris, 91191 Gif-sur-Yvette, France\\
	$^{77}$ Centro de Investigaciones Energ\'eticas, Medioambientales y Tecnol\'ogicas (CIEMAT), Avenida Complutense 40, 28040 Madrid, Spain\\
	$^{78}$ Instituto de Astrof\'isica e Ci\^encias do Espa\c{c}o, Faculdade de Ci\^encias, Universidade de Lisboa, Tapada da Ajuda, 1349-018 Lisboa, Portugal\\
	$^{79}$ Universidad Polit\'ecnica de Cartagena, Departamento de Electr\'onica y Tecnolog\'ia de Computadoras,  Plaza del Hospital 1, 30202 Cartagena, Spain\\
	$^{80}$ Institut de Recherche en Astrophysique et Plan\'etologie (IRAP), Universit\'e de Toulouse, CNRS, UPS, CNES, 14 Av. Edouard Belin, 31400 Toulouse, France\\
	$^{81}$ Kapteyn Astronomical Institute, University of Groningen, PO Box 800, 9700 AV Groningen, The Netherlands\\
	$^{82}$ INAF-Osservatorio Astronomico di Brera, Via Brera 28, 20122 Milano, Italy\\
	$^{83}$ INAF-Istituto di Astrofisica e Planetologia Spaziali, via del Fosso del Cavaliere, 100, 00100 Roma, Italy\\
	$^{84}$ Dipartimento di Fisica "Aldo Pontremoli", Universit\'a degli Studi di Milano, Via Celoria 16, 20133 Milano, Italy\\
	$^{85}$ INFN-Sezione di Milano, Via Celoria 16, 20133 Milano, Italy\\
	$^{86}$ Dipartimento di Fisica e Astronomia "Augusto Righi" - Alma Mater Studiorum Universit\'a di Bologna, Viale Berti Pichat 6/2, 40127 Bologna, Italy\\
	$^{87}$ Junia, EPA department, 41 Bd Vauban, 59800 Lille, France\\
	$^{88}$ SISSA, International School for Advanced Studies, Via Bonomea 265, 34136 Trieste TS, Italy\\
	$^{89}$ IFPU, Institute for Fundamental Physics of the Universe, via Beirut 2, 34151 Trieste, Italy\\
	$^{90}$ INFN, Sezione di Trieste, Via Valerio 2, 34127 Trieste TS, Italy\\
	$^{91}$ Dipartimento di Fisica e Scienze della Terra, Universit\'a degli Studi di Ferrara, Via Giuseppe Saragat 1, 44122 Ferrara, Italy\\
	$^{92}$ Istituto Nazionale di Fisica Nucleare, Sezione di Ferrara, Via Giuseppe Saragat 1, 44122 Ferrara, Italy\\
	$^{93}$ Dipartimento di Fisica - Sezione di Astronomia, Universit\'a di Trieste, Via Tiepolo 11, 34131 Trieste, Italy\\
	$^{94}$ NASA Ames Research Center, Moffett Field, CA 94035, USA\\
	$^{95}$ INAF, Istituto di Radioastronomia, Via Piero Gobetti 101, 40129 Bologna, Italy\\
	$^{96}$ INFN-Bologna, Via Irnerio 46, 40126 Bologna, Italy\\
	$^{97}$ Institute for Theoretical Particle Physics and Cosmology (TTK), RWTH Aachen University, 52056 Aachen, Germany\\
	$^{98}$ Institute for Astronomy, University of Hawaii, 2680 Woodlawn Drive, Honolulu, HI 96822, USA\\
	$^{99}$ Department of Physics \& Astronomy, University of California Irvine, Irvine CA 92697, USA\\
	$^{100}$ University of Lyon, UCB Lyon 1, CNRS/IN2P3, IUF, IP2I Lyon, 4 rue Enrico Fermi, 69622 Villeurbanne, France\\
	$^{101}$ INFN-Sezione di Genova, Via Dodecaneso 33, 16146, Genova, Italy\\
	$^{102}$ School of Physics, HH Wills Physics Laboratory, University of Bristol, Tyndall Avenue, Bristol, BS8 1TL, UK\\
	$^{103}$ Ruhr University Bochum, Faculty of Physics and Astronomy, Astronomical Institute (AIRUB), German Centre for Cosmological Lensing (GCCL), 44780 Bochum, Germany\\
	$^{104}$ Department of Physics, Lancaster University, Lancaster, LA1 4YB, UK\\
	$^{105}$ Univ. Grenoble Alpes, CNRS, Grenoble INP, LPSC-IN2P3, 53, Avenue des Martyrs, 38000, Grenoble, France\\
	$^{106}$ Department of Physics and Astronomy, University College London, Gower Street, London WC1E 6BT, UK\\
	$^{107}$ Department of Physics and Helsinki Institute of Physics, Gustaf H\"allstr\"omin katu 2, 00014 University of Helsinki, Finland\\
	$^{108}$ Astrophysics Group, Blackett Laboratory, Imperial College London, London SW7 2AZ, UK\\
	$^{109}$ Dipartimento di Fisica, Sapienza Universit\`a di Roma, Piazzale Aldo Moro 2, 00185 Roma, Italy\\
	$^{110}$ Department of Mathematics and Physics E. De Giorgi, University of Salento, Via per Arnesano, CP-I93, 73100, Lecce, Italy\\
	$^{111}$ INFN, Sezione di Lecce, Via per Arnesano, CP-193, 73100, Lecce, Italy\\
	$^{112}$ INAF-Sezione di Lecce, c/o Dipartimento Matematica e Fisica, Via per Arnesano, 73100, Lecce, Italy\\
	$^{113}$ Institute for Computational Science, University of Zurich, Winterthurerstrasse 190, 8057 Zurich, Switzerland\\
	$^{114}$ Higgs Centre for Theoretical Physics, School of Physics and Astronomy, The University of Edinburgh, Edinburgh EH9 3FD, UK\\
	$^{115}$ Institut f\"ur Theoretische Physik, University of Heidelberg, Philosophenweg 16, 69120 Heidelberg, Germany\\
	$^{116}$ Universit\'e St Joseph; Faculty of Sciences, Beirut, Lebanon\\
	$^{117}$ Department of Astrophysical Sciences, Peyton Hall, Princeton University, Princeton, NJ 08544, USA\\
	$^{118}$ Department of Astronomy, University of Massachusetts, Amherst, MA 01003, USA}

   \date{Received 25 May 2023 / Accepted 26 September 2023}

 
  \abstract{The technique of photometric redshifts has become essential for the exploitation of multi-band extragalactic surveys. While the requirements on \phz s for the study of galaxy evolution mostly pertain to the precision and to the fraction of outliers, the most stringent requirement {in} their use in cosmology is on the accuracy, with a level of bias at the sub-percent level for the \Euclid cosmology mission. A separate, and challenging, calibration process is needed to control the bias at this level of accuracy. {The bias in \phz s has several distinct origins that may not always be easily overcome.} We identify here one source of bias linked to the spatial or time variability of the passbands used to determine the photometric colours of galaxies. We first quantified the effect as observed on several well-known photometric cameras, and {found in particular that, due to the properties of optical filters, the redshifts of off-axis sources are usually overestimated}. We show using simple simulations that the detailed and complex changes in the shape can be mostly ignored and that it is sufficient to know the {mean} wavelength of the passbands of each photometric observation to correct almost exactly for this bias; {the key point is that this mean wavelength is independent of the spectral energy distribution of the source}. We use this property to propose a {correction that can be computationally efficiently implemented in some \phdz\ algorithms}, in particular template-fitting. We verified that our algorithm, implemented in the new \phdz\ code \texttt{Phosphoros}, can effectively reduce the bias in \phz s on real data using the CFHTLS T007 survey, with an average measured bias $\Delta z$ over the redshift range $0.4\le z\le 0.7$ decreasing by about 0.02, {specifically from $\Delta z\simeq 0.04$ to $\Delta z\simeq 0.02$ around $z=0.5$}. Our algorithm is also able to produce corrected photometry for other applications.}

   \keywords{Galaxies: distances and redshifts --
   	Cosmology: observations --
   	Surveys --
   	Techniques: photometric --
    Techniques: miscellaneous}
   
   \titlerunning{Variations of photometric passbands and photometric-redshift accuracy}
   \authorrunning{Euclid Collaboration}

   \maketitle
%

\section{Introduction}
{Multi-megapixel cameras with large fields of view have revolutionised extragalactic astrophysics and observational cosmology by enabling photometric surveys of large sky areas in several optical and near-infrared bands}. The Dark Energy Survey \citep[DES;][]{TheDarkEnergySurveyCollaboration2005}, the Kilo-Degree Survey \citep[KiDS;][]{deJong2013}, {and} the Hyper Suprime-Cam Strategic Survey Program \citep[HSC-SSP;][]{Aihara2018} are recent examples of photometric surveys with areas exceeding 1000\,deg$^2$. These surveys enable the measurement of {the} cosmic shear, which is the distortion of the images of distant objects caused by the propagation of light rays through inhomogeneous matter \citep{Blandford1991}. The cosmic shear allows the reconstruction of dark-matter maps at different redshifts, from which the distribution of matter and its evolution can be inferred. {Modern} cosmological surveys have established cosmic shear as one of the main modern cosmological probes and have already provided important cosmological constraints {(e.g. \citealt{Abbott2018} for DES; \citealt{Asgari2021} for KiDS;  and \citealt{Hamana2020} for HSC-SSP)}. 

\Euclid \citep{Laureijs2011} is a mission of the European Space Agency that will perform a survey {over 15\,000\,deg$^{2}$ of extragalactic sky} \citep{Scaramella2022} with optical and near-infrared imaging, as well as with slitless multi-object spectroscopy in the near-infrared. The main scientific probes of \Euclid are the cosmic shear and {galaxy clustering}. They are supported by two instruments: The VIS optical camera \citep{Cropper2014} will provide us with high-resolution images of galaxies for the determination of the cosmic shear; and the Near Infrared Spectrometer and Photometer  near-infrared instrument \citep[NISP;][]{Maciaszek2016} will perform near-infrared photometry in three bands to support cosmic shear determination, as well as near-infrared spectroscopy for the study of {three-dimensional galaxy clustering}. Compared to current ground-based surveys, the determination of the cosmic shear with \Euclid will greatly benefit from high-resolution imaging and {near-infrared photometry} from space, but also from the significantly larger sky area (three times the area of DES) and {depth}.

The measurement of redshifts for a large fraction of the surveyed galaxies is an indispensable step in the cosmological study of the cosmic shear. While redshift measurements can be performed through spectroscopy, spectroscopy is more challenging than photometry at very faint fluxes, which limits the number of measurable redshifts, even when efficient multi-object spectrographs are available. In the case of \Euclid, spectroscopic redshifts will be determined for some 30 million galaxies, while the total number of galaxies for which sufficiently precise shapes can be measured {will} exceed one billion \citep{Laureijs2011}. {The technique of \phz s, that is, using photometric observations only, is currently the only way to determine redshifts on such a huge scale}. {The limited precision of \phz s is not an issue for cosmic shear because the efficiency of lensing is a slowly varying function of the redshift}. Photometric redshifts have thus become an essential tool of modern observational cosmology.

Photometric-redshift determination requires photometric observations of galaxies in several wavelength bands, which defines a multi-dimensional flux or colour space\footnote{Here we consider colour spaces only.}. The redshift is then obtained from the construction of a mapping between the position of an object in this colour space and the redshift. We can in principle determine this mapping through different approaches, either based on real objects with known redshifts or on simulated objects. The determination of redshifts of galaxies using a small number of broad-band photometric measurements {was} pioneered six decades ago by \citet{Baum1962}. The template-fitting (TF) technique, which involves the comparison of the magnitudes or fluxes of a galaxy with those of simulated objects, {was} first developed by \citet{Loh1986} and subsequently {exploited} in several codes that are still in use today {\citep[e.g.][]{Arnouts1999,Bolzonella2000}}. A more recent approach using machine-learning (ML), {whose first applications to \phdz\ calculations were made by \citet{Firth2003} and \citet{Tagliaferri2003} using artificial neural networks}, is now becoming more and more popular, and virtually every ML approach has been investigated, for instance support vector machine \citep{Wadadekar2005}, decision trees \citep{CarrascoKind2013}, and Gaussian processes \citep{Almosallam2016}. A review of the challenges of \phdz\ determination in the context of cosmological surveys has been presented in \citet{Newman2022}.

The usefulness of any quantity for a scientific application is bound to meet {the} requirements on the quality of its determination. In the case of \phz s, this quality is usually expressed with three parameters. When a \phdz\ determination is successful, the measurement is distributed around the true value; the dispersion $\sigma_z$ of this distribution is a measure of the precision of the determination. However, it sometimes happens that the \phdz\ determination fails completely and the predicted redshift is found to lie very far from the true value; the probability of such a failure is called the `outlier fraction'. Finally, the bias $\Delta z$ is the location of the peak of the distribution of the differences between predictions and true redshift values and determines the accuracy of the predictions.  

The \Euclid requirements on the precision of \phz s {\citep{Amara2007}} are not extremely demanding [$\sigma_z=0.05(1+z)$ and an outlier fraction of $10$\% at a magnitude 24.5] compared {with} what can be achieved on small fields{; for instance, in the COSMOS field, \citet{Weaver2022} obtained, at similar depths, $\sigma_z$ better than $0.02(1+z)$ and an outlier fraction better than $5$\%. Meeting the \Euclid requirements nevertheless} remains challenging over the full Euclid wide survey because of the difficulty to obtain deep-enough photometry over a field almost 10\,000 times larger \citep{Desprez2020}. However, the biggest difficulty in using \phz s for cosmology is the accuracy in the \phdz\ determination that is necessary for the cosmic-shear probe; in \Euclid, the requirement is expressed as the bias on the mean redshift in each of the approximately ten tomographic redshift bins at redshift $z$, which needs to be less than $\Delta z/(1+z)=0.002$. Such {a} stringent requirement demands that some bias correction takes place after the \phdz\ determination, for instance by calibrating the bias directly in the colour space occupied by galaxies \citep{Masters2015}.{ This study was the main motivation behind the C3R2 project to gather a gold-standard set of spectroscopic redshifts over the full colour space of galaxies \citep{Masters2017,Masters2019,Guglielmo2020,Stanford2021,Saglia2022}}. It is however essential to remove, as much as possible, any bias before the calibration step if we want to meet this requirement. We focus here on one specific source of bias that is inherent to the photometric measurement and that is linked to the time and spatial variations of the photometric passbands. {Spatial variations are dependent on the positions of the sources in the field of view and thus induce spatial variations {unrelated to the source} in the \phz s that, in turn, introduce a spurious signal into the correlation function of the cosmic shear}. {Such variations may have an effect on the dispersion, outlier fraction, and bias; we focus here on the bias since this is the most stringent requirement on \phz s for \Euclid}. 

{In this paper, we first review the current knowledge about passband variations for surveys relevant to \Euclid. We then discuss the problems caused by these variations and demonstrate the effect quantitatively using idealised simulations. We finally propose an efficient implementation that can remove most of the bias resulting from this issue, and validate it using a real photometric survey. While we focus on the spatial dependence in this paper, the correction also applies to time variations of the passbands.}

\section{Photometry}
Photometric measurements are performed over ranges of wavelengths called `passbands' that are defined by a transmission curve. This curve, denoted $T(\lambda)$, is a function of the wavelength $\lambda$ that describes the fraction of photons (or of the energy) {entering the telescope} that is ultimately recorded on the detector. {The main component affecting the transmission curve is an optical filter that restricts the range of accessible wavelengths; other components include the detector quantum efficiency and the transmission through the other optical elements}.  Usually $T(\lambda)$ transmissions are designed to approximate more or less a top-hat filter. One defines the AB flux $F_T$ of a source with {rest-frame} spectral energy distribution (SED) $L(\lambda_0)$, with $\lambda_0=\lambda/(1+z)$ where $z$ is the redshift, through the photometric transmission curve $T(\lambda)$ as \citep{Oke1983}
\begin{equation}
\label{eq:flux_ab}
F_T = \frac{\int_0^\infty s(\lambda)\, T(\lambda)\, \lambda\, \mathrm{d}\lambda}{\mathrm{c}\int_0^\infty T(\lambda)\,\frac{\dd\lambda}{\lambda}}\;,
\end{equation}
where $c$ is the speed of light and
\begin{equation}
\label{eq:flux_lum}
s(\lambda)=\frac{L(\lambda_0)\,10^{-0.4E_{B-V}k_\lambda}}{4\pi\,(1+z)\,D_\mathrm{L}^2}
\end{equation}
is the observed flux as a function of wavelength, with $D_\mathrm{L}$ the luminosity distance, $E_{B-V}$ the Galactic reddening and $k(\lambda)$ the extinction law. The above expression assumes that the detector is counting photons, which is the case for the vast majority of modern detectors in the optical and near-infrared range. The AB convention provides the flux of a source with flat SED (when expressed per unit of frequency) that would leave the same number of counts on the detector as the source under study when passing through the transmission curve $T(\lambda)$. We point out that, with the definition of Eq.~\eqref{eq:flux_ab}, the normalisation of $T(\lambda)$ is arbitrary. {We assume here that $T(\lambda)$ is normalised so that $\int_0^\infty T(\lambda)\,\dd\lambda=1$}.

\begin{figure*}
	\includegraphics[width=\textwidth]{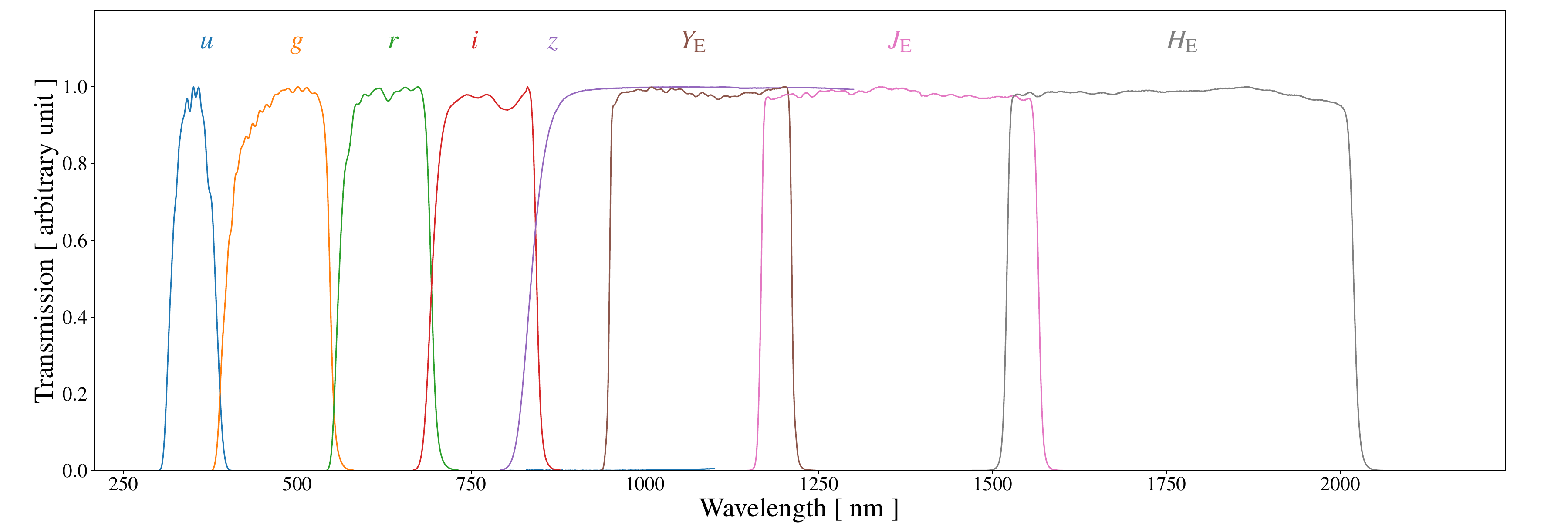}\\
	\caption{\label{fig:ugrizYJH}Set of transmission curves ${\cal T}=\{ugriz$\YE\JE\HE\} used for the \Euclid mission (from left to right). The $ugriz$ passbands are only fiducial, since different sets will be used by \Euclid; those represented here are from SDSS. The \YE\JE\HE\ passbands are from NISP on board \Euclid. {Only the filter transmissions are shown, without atmospheric, telescope, and detector quantum efficiency effects.}}
\end{figure*}

Photometric extragalactic surveys use a (small) set of $N$ passbands with transmission curves $T_i(\lambda)$, $i=1,\ldots,N$, covering distinct, and usually very slightly overlapping, wavelength ranges.
The $ugriz$ set of passbands of the Sloan Digital Sky Survey Photometric Camera \citep{Gunn1998} is now the most common system used for extragalactic surveys, with slight differences in the filters depending on the cameras and telescopes. The \Euclid photometric survey consists of the very broad \IE\ passband of the VIS optical instrument, which covers roughly the $riz$ bands and which we will not consider further here, and the three passbands in the near-infrared of the NISP instrument, the \YE, \JE, and \HE\ passbands (see Sect.~\ref{sec:nisp}). The full set of passbands used in this paper, which we denote $\cal T$, are shown in Fig.~\ref{fig:ugrizYJH}.

\section{Spatial variation of photometric passbands}
A consistent photometric system implicitly assumes that the transmissions $T_i(\lambda)$, $i=1,\ldots,N$ are the same for all objects. However, small variations from object to object are possible. In the presence of passband variations, the colours of the different sources occupy different colour spaces, and cannot be compared any more. Passband variations can occur due to several effects as a function of time or of position on the detector. Time dependence can be introduced {by atmospheric effects} and by the evolution or degradation of the properties of the optical elements, filters, or detectors in a way that depends on the wavelength. Spatial dependence of the passbands is another issue that affects differently, and systematically, the sources in the field of view. Such effects can be introduced by non-uniformities of the filters or detectors, or by the fact that the optical beams of off-axis sources hit the filter at angles that are different from those in the case of on-axis sources. The latter effect can in principle be predicted. In order to obtain photometric measurements with the highest possible accuracy, several teams have measured passband variations across the field of view of their camera. {We briefly review {studies of the spatial variations of the passbands} below, in order to demonstrate that the issue is general, and not limited to a particular survey.}

\label{sec:measurement}

{Passband variations have been measured for several photometric systems}. Denoting $E(x)$ the average of {any function} $x(\lambda)$ over the {transmission} $T(\lambda)$, that is,
\begin{equation}
	E(x)=\frac{\int_0^\infty x(\lambda)\,T(\lambda)\,\dd\lambda}{\int_0^\infty T(\lambda)\,\dd\lambda},
\end{equation}
we characterise the measured variations using the changes in the first four moments of the normalised transmission curve $T(\lambda)$ in different locations in the field of view: the mean $\mu=E(\lambda)$; the variance $\sigma^2=E\left((\lambda-\mu)^2\right)$; the skewness $\Sigma=E\left(\left(\frac{\lambda-\mu}{\sigma}\right)^3\right)$; and the kurtosis $\kappa=E\left(\left(\frac{\lambda-\mu}{\sigma}\right)^4\right)$. We point out that the skewness and the kurtosis are the standardised  third and fourth moments, respectively. We will also use the dispersion $\sigma$, instead of the second moment $\sigma^2$, as it is more intuitively understandable. {We stress that these moments do not depend on the SED of the source}.

\subsection{Sloan Digital Sky Survey Photometric Camera}
\begin{figure}
	\includegraphics[width=8.8cm]{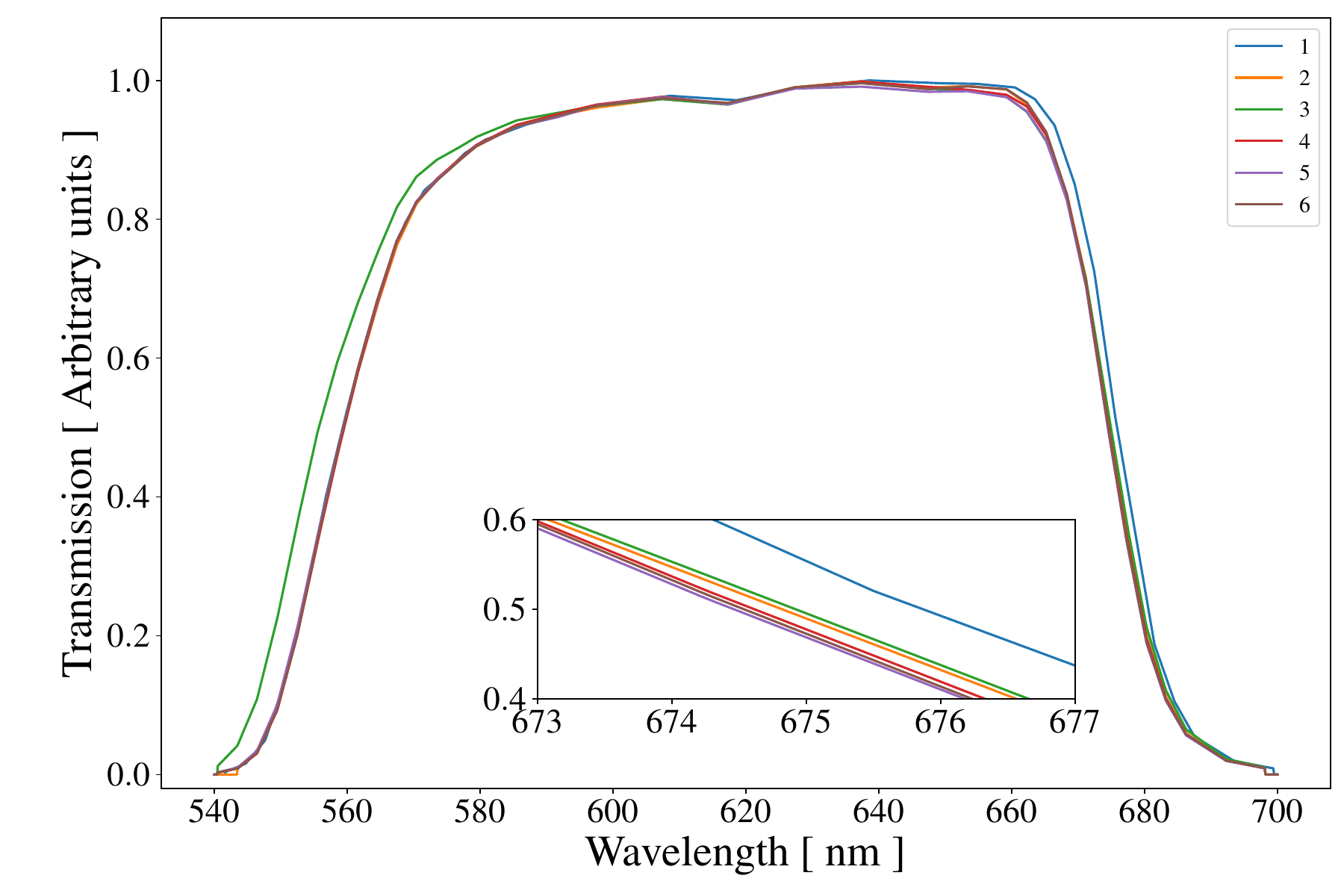}\\
	\includegraphics[width=8.8cm]{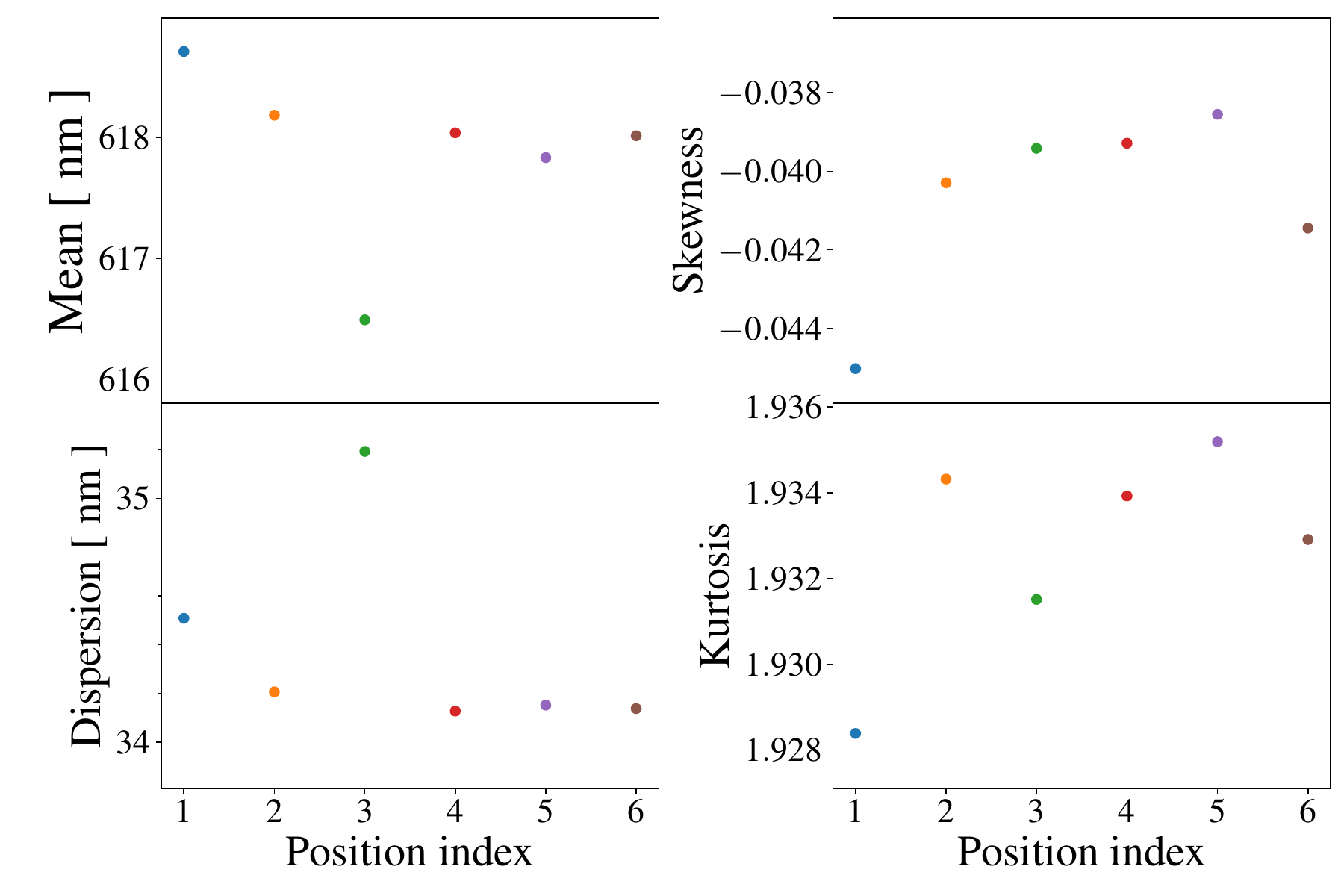}
	\caption{\label{fig:sdss}Passband variations of the SDSS $r$ filter. \emph{Top}: Measured variations of the SDSS $r$ filters from \citet{Doi2010}. Measurements have been performed for each of the six different CCD columns. {The transmissions have renormalised so that the maximum of the transmission {at Position index 1} is 1.} {The inset shows a zoom on the cut-off of the transmissions. The legend gives the position indices associated to each colour.} \emph{Bottom}: The first four moments of the six SDSS $r$ passbands. {Each transmission is identified with a specific colour, identical in the top and bottom panels.}}
\end{figure}

The response of the Sloan Digital Sky Survey Photometric Camera \citep{Gunn1998}, which is mounted on the Sloan Foundation Telescope at Apache Point Observatory, has been characterised in extensive {detail} in \citet{Doi2010}. {The camera is composed of 30 CCDs in a matrix of 5 rows and 6 columns.} The passband variations are determined empirically using a series of monochromatic 1\,nm-wide dome flats \citep{Rheault2012}. They measured in particular the response for the six columns of the imager for the five $ugriz$ bands, in order to estimate the column-to-column passband variation. Figure~\ref{fig:sdss} {(top)} shows the six $r$ passbands obtained in one campaign\footnote{The data for all filters can be found at \url{http://www.ioa.s.u-tokyo.ac.jp/~doi/sdss/SDSSresponse.html}}. Stronger variations due to temperature and ageing are however reported by \citet{Doi2010}; they quote variations in the passband effective wavelengths from about 3\,nm in the best case ($g$ and $i$ passbands) to 12\,nm in the worst case ($z$ passband, which is strongly affected by the CCD quantum efficiency), and 3.5\,nm in the $r$ passband. Unfortunately, the measurements obtained at other periods do not seem {to be} available.

Figure~\ref{fig:sdss} {(bottom)} shows the first four moments of the $r$ passbands. One {detector} column (3) shows particularly large variations of mean wavelength and dispersion. The overall shape remains however very similar, with very {minor} {change} in skewness ($< 0.01$) and kurtosis (0.007).

\subsection{Dark Energy Camera}
\begin{figure}
	\includegraphics[width=8.8cm]{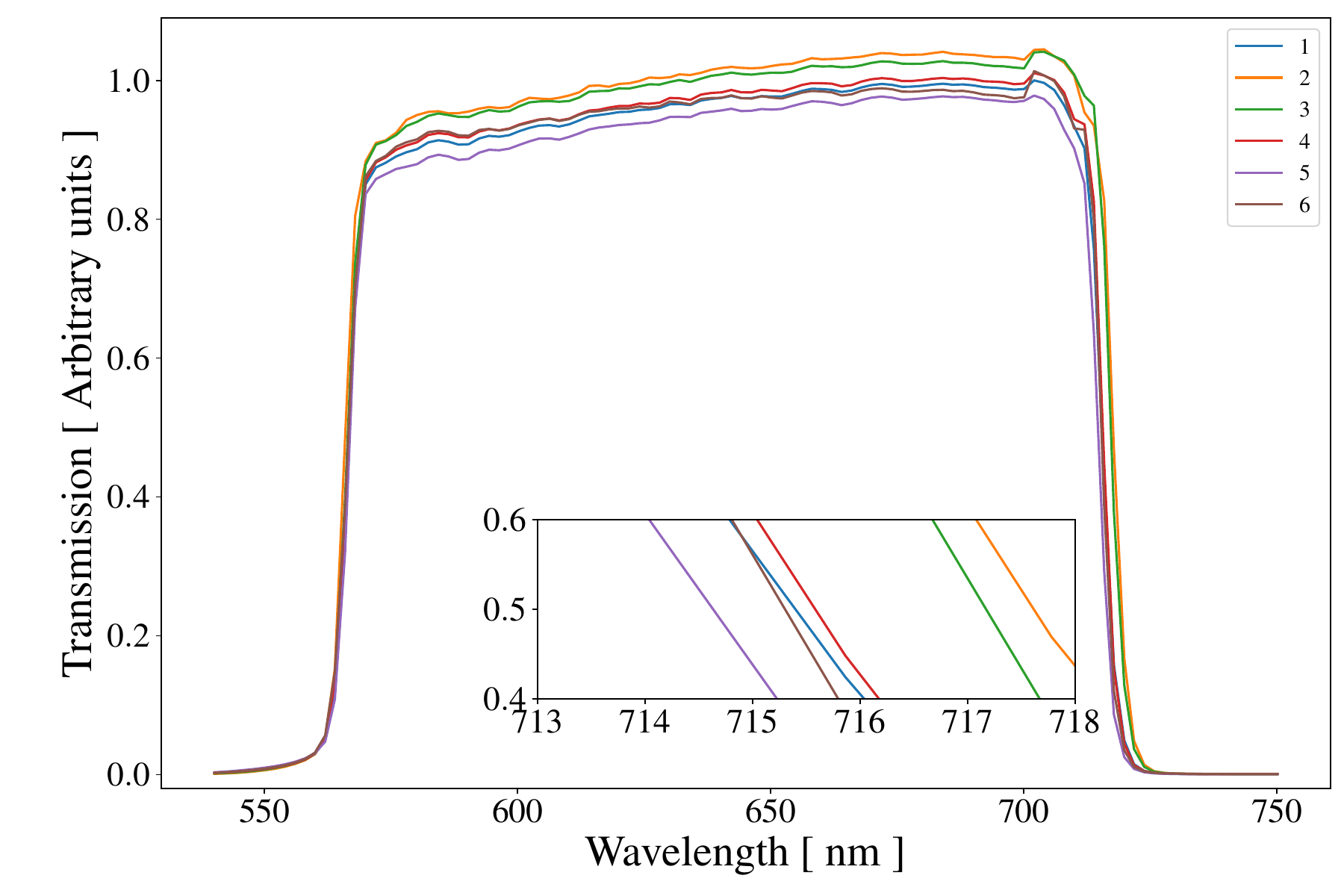}\\
	\includegraphics[width=8.8cm]{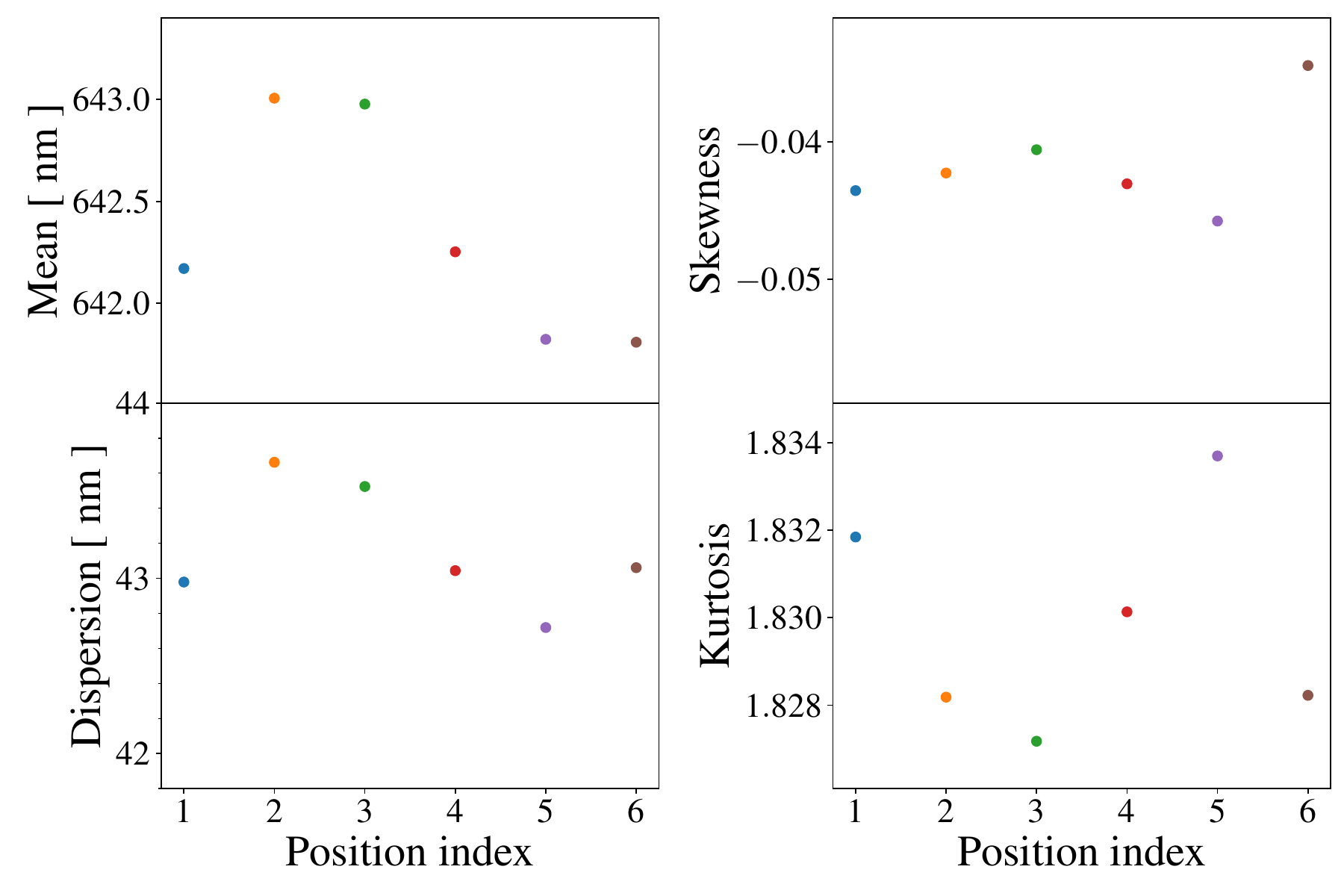}
	\caption{\label{fig:des}Passband variations of the DES $r$ filter. \emph{Top}: Measured variations of the DES $r$ filters from \citet{Li2016}. Six measurements have been performed at different off-axis positions; {larger position indices indicate larger off-axis distances}. {The transmissions have renormalised so that the maximum of the transmission {at Position index 1} is 1.} {The inset shows a zoom on the cut-off of the transmissions. The legend gives the position indices associated to each colour.} \emph{Bottom}: The first four moments of the six DES $r$ passbands. {Each transmission is identified with a specific colour, identical in the top and bottom panels.}}
\end{figure}

The Dark Energy Survey \citep{TheDarkEnergySurveyCollaboration2005} is a $griz$ photometric survey of about 5000\,deg$^2$ with the 2.2\,deg-diameter Dark Energy Camera \citep{Flaugher2015} located on the Blanco telescope at Cerro Tololo Observatory. Passband variations have been studied in detail by \citet{Li2016}. {They measured} the change of the passband as a function of the distance to the centre of the detector {up to the edge of the field-of-view around 1.1 deg}. The $i$ filter was found to {show} the largest {variation}, with a wavelength shift of the cut-on wavelength of about 6\,nm. Figure~\ref{fig:des} {(top)} shows the variation of the $r$ passband. As in the case of SDSS, there is little variation in the shape of the passband. 

Figure~\ref{fig:des} {(bottom)} shows the first four moments of the $r$ passbands; {larger position indices indicate larger off-axis distances}. The mean wavelength is very stable, with an amplitude of variation of about 1\,nm {compared to the passband in the centre of the field of view, which we refer to in the following as the `central passband', without clear dependence on the off-axis angle}; the dispersion variation is very comparable to that of SDSS. Again very small changes in skewness and kurtosis are observed.

\subsection{MegaCam}
\label{sec:megacam}
\begin{figure}
\includegraphics[width=8.8cm]{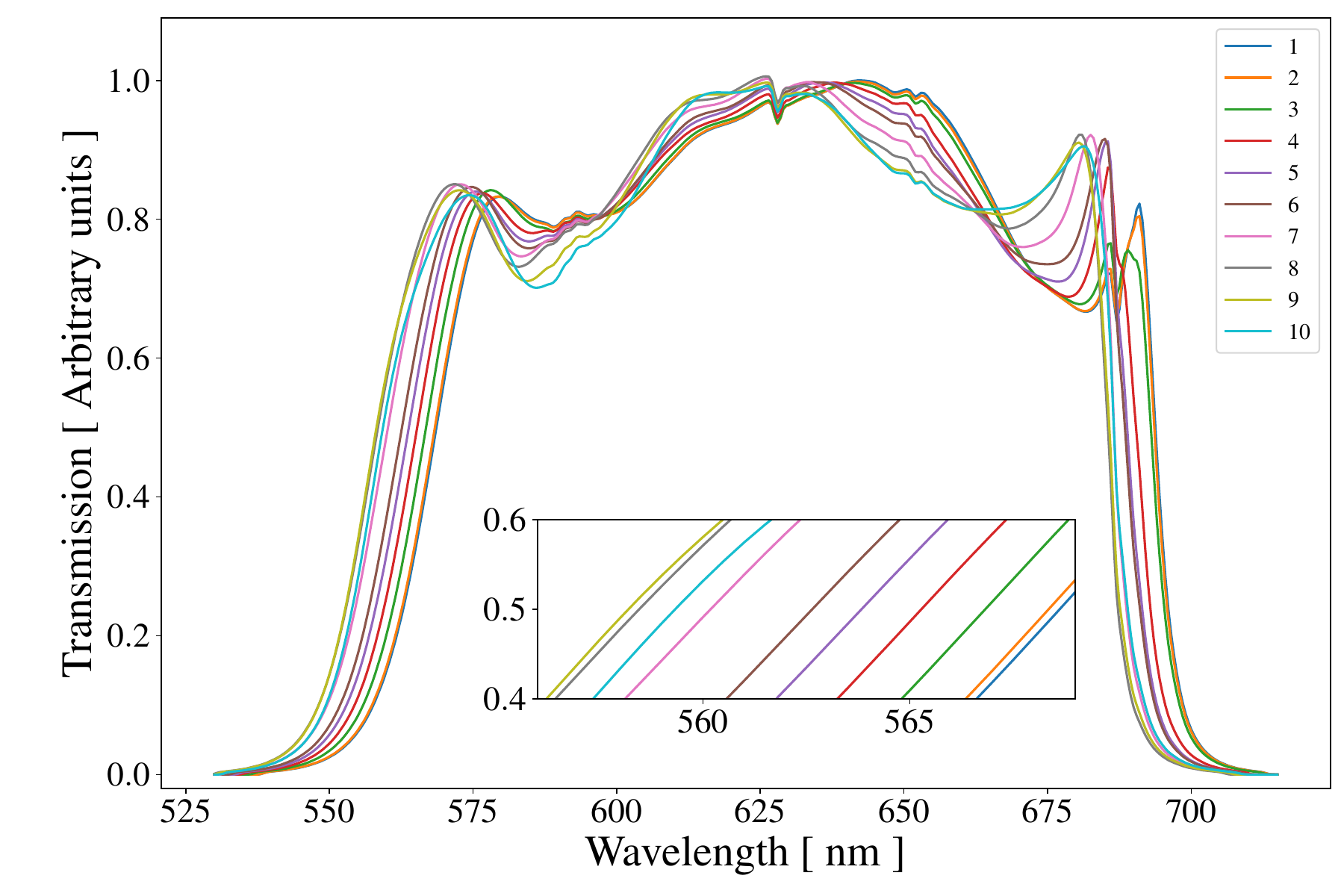}\\
\includegraphics[width=8.8cm]{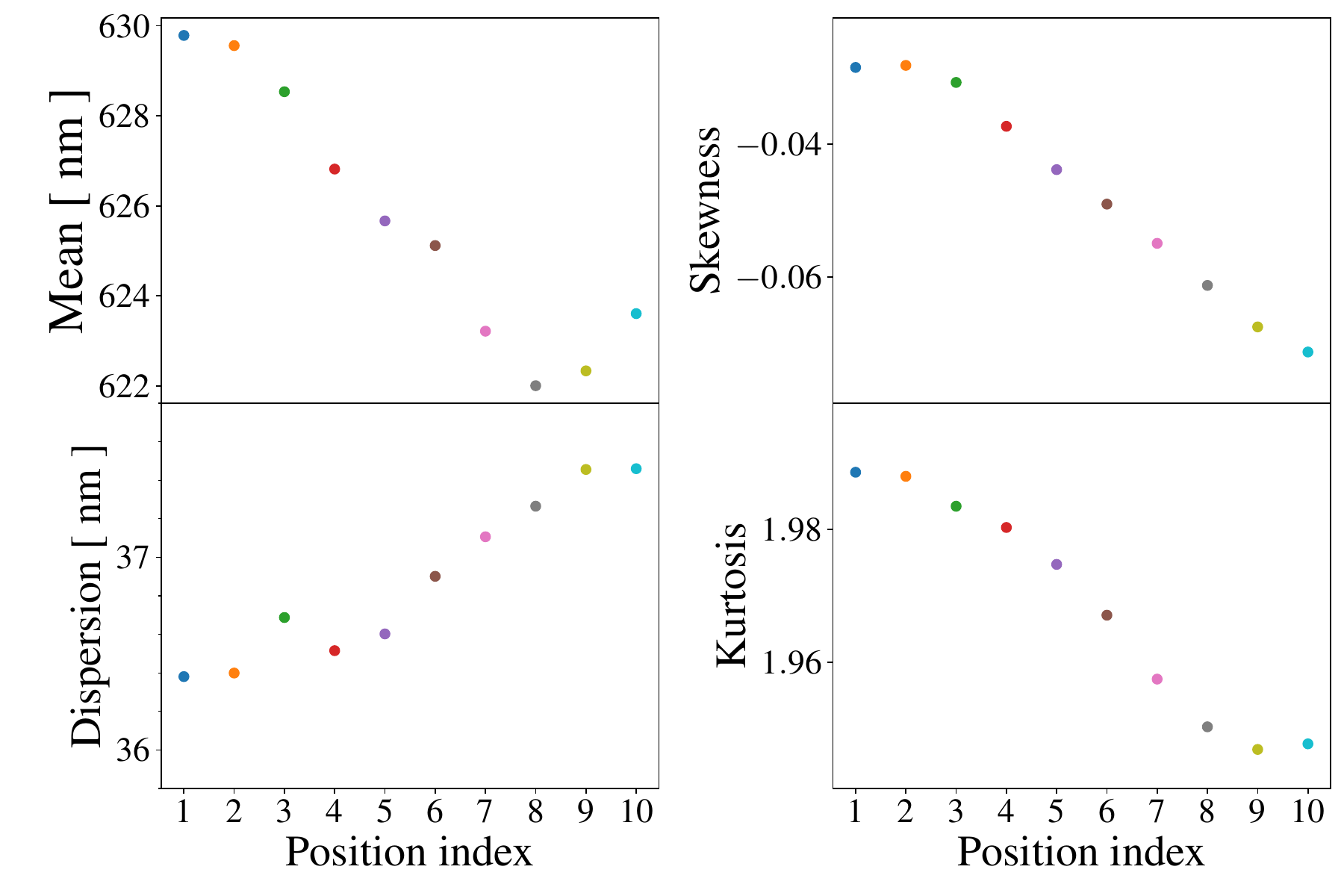}
\caption{\label{fig:megacam}Passband variations of the MegaCam $r$ filter. \emph{Top}: Measured variations of the MegaCam $r$ filters from \citet{Betoule2013}. Ten measurements have been performed at different off-axis positions; {larger position indices indicate larger off-axis distances}. {The transmissions have renormalised so that the maximum of the transmission {at Position index 1} is 1.} {The inset shows a zoom on the cut-on of the transmissions. The legend gives the position indices associated to each colour.} \emph{Bottom}: The first four moments of the ten MegaCam $r$ passbands. {Each transmission is identified with a specific colour, identical in the top and bottom panels.}}
\end{figure}

The MegaCam instrument \citep{Boulade2003} is a 1-deg$^2$ imaging camera located on the prime focus of the Canada-France-Hawaii Telescope in Hawaii. The detailed calibration of the camera has been performed in the framework of the Supernova Legacy Survey project \citep{Guy2010} by \citet{Betoule2013}. They found the location of the source as a function of the distance to the centre of the detector plane to be a major driver of passband variations{, with transmissions moving towards shorter wavelengths. This is expected for optical interference filters \citep[see discussion in, e.g., section 3.2 of][and references therein]{Schirmer2022}}. Figure~\ref{fig:megacam} {(top)} shows the variation of the $r$ passband, where the position index is correlated with the off-axis distance, {with every shift in the position index corresponding to about an additional 5 arcmin distance from the centre of the field (See also Sect.~\ref{sec:appli})}.

Figure~\ref{fig:megacam} {(bottom)} shows the first four moments of the $r$ passbands. The changes in the mean wavelength are much more significant, with a peak-to-peak amplitude of about 8\,nm, although the change in the dispersion is not significantly {larger} than in the case of SDSS or DES. Similar changes in mean wavelengths do occur with other SDSS or DES passbands. The amplitude of skewness and (especially) kurtosis variations are several times larger than for SDSS, indicating the presence of more significant changes in the shape of the passband.

\subsection{Euclid NISP}
\label{sec:nisp}
\Euclid NISP \citep{Maciaszek2016} is equipped with 3 photometric filters covering the \Euclid \YE\JE\HE\ bands, shown in Fig.~\ref{fig:ugrizYJH}. The NISP photometric system is described in detail in \citet{Schirmer2022}. The passbands were computed as a function of position in the NISP field of view, based on local filter passband measurements and full ray tracing of the \Euclid NISP optical system to account for angle-of-incidence variations on the filter surface, {reaching an accuracy of 0.8 nm}. {The study focuses on the determination of cut-on and cut-off wavelengths, but does not addresses other variations of the transmission}. Polynomial expressions have been provided to compute {the cut-on and cut-off wavelengths} at any position \citep[see figure 8 in][]{Schirmer2022}. These measurements showed the existence of a blue shift that depends on the off-axis distance, which can be explained by the different incident angles of the incoming light. Blue shifts between 2.5\,nm and 6.1\,nm have been observed for the cut-on or cut-off wavelengths of the three filters, with the \YE\ and \HE\ filters being the least and the most affected, respectively \citep[see figure 9 in][]{Schirmer2022}.  

\section{Consequences of passband variations}
\subsection{Effect on the photometry}
\label{sec:photom}
{Photometric observations require a calibration in order to derive, for each observation frame, the so-called zero-point, i.e. the relation between the physical flux (or magnitude) and the count rate on the detector. This is generally achieved by using reference stars located in the field of view. The AB definition from Eq.~\eqref{eq:flux_ab} provides an exact relation only for sources that have the same SED as the calibration stars. In order to cope with galaxies with different SEDs, the calibration is sometimes refined with the addition of a colour term \citep[e.g.,][]{Padmanabhan2008,deJong2015}, which is a correction based on the ratio of the counts in adjacent photometric bands, providing a coarse approximation for the true shape of $s(\lambda)$. This method calibrates all bands simultaneously, which can be done only for the bands that are observed with the same telescope. It is therefore not well adapted to the \Euclid survey, which will use a combination of telescopes, and thus colour-term correction is not used.}

{With regard to passband variations, the fact that a unique zero-point correction is computed for each frame implies that any spatial variations of the passband across the field-of-view are ignored. The calibration is therefore formally valid only for some average passband.}  {In the case of \Euclid, a second step is performed which corrects} for any spatially dependent systematic deviations of the reconstructed fluxes. This step makes the fluxes independent of possible changes in the normalisation of the passband, provided they only depend on the location in the field of view. {In the case of passband variations, this correction would remove any effect linked to a change in the effective area of the transmission.} {Atmospheric absorption also results in passband variations; however, they can be considered as spatially uniform over a single observation. Hence the calibration process will absorb this effect in the zero point, although in reality the correction depends on the colour of the object.} {For extended objects, the passband could in principle change across the object, which would affect photometric extraction in a complicated way; however, in cosmological applications useful galaxies are very small compared to the scale on which passband variations are measured, so we can safely ignore this effect.}

As a result of the \Euclid photometric calibration, variations that are not limited to a change in normalisation do impact the photometric measurements. In their very detailed analysis of the photometric stability of the SDSS camera, \citet{Doi2010} found that column-to-column variations of the passbands induce errors on the $g$, $r$, $i$ and $z$ fluxes of up to 1\% (0.01 mag). {Based on the `Scientific Challenge 8' simulations of the \Euclid performance, \citet{Schirmer2022} found that the effect is of the order of a few millimags in the \Euclid near-infrared bands}. Such bias would impact the performance of \phdz\ determination. In order to remove this bias, it would be necessary to include the correct transmission curve $T(\lambda)$ in Eq.~\eqref{eq:flux_ab}; however, the bias would also depend on the a priori unknown SED of the object, so that the correction cannot be performed on isolated frames.

\subsection{Effects on \phdz\ determination}
\label{sec:algo}

\subsubsection{First-order effect}
\begin{figure}
	\includegraphics[width=8.8cm]{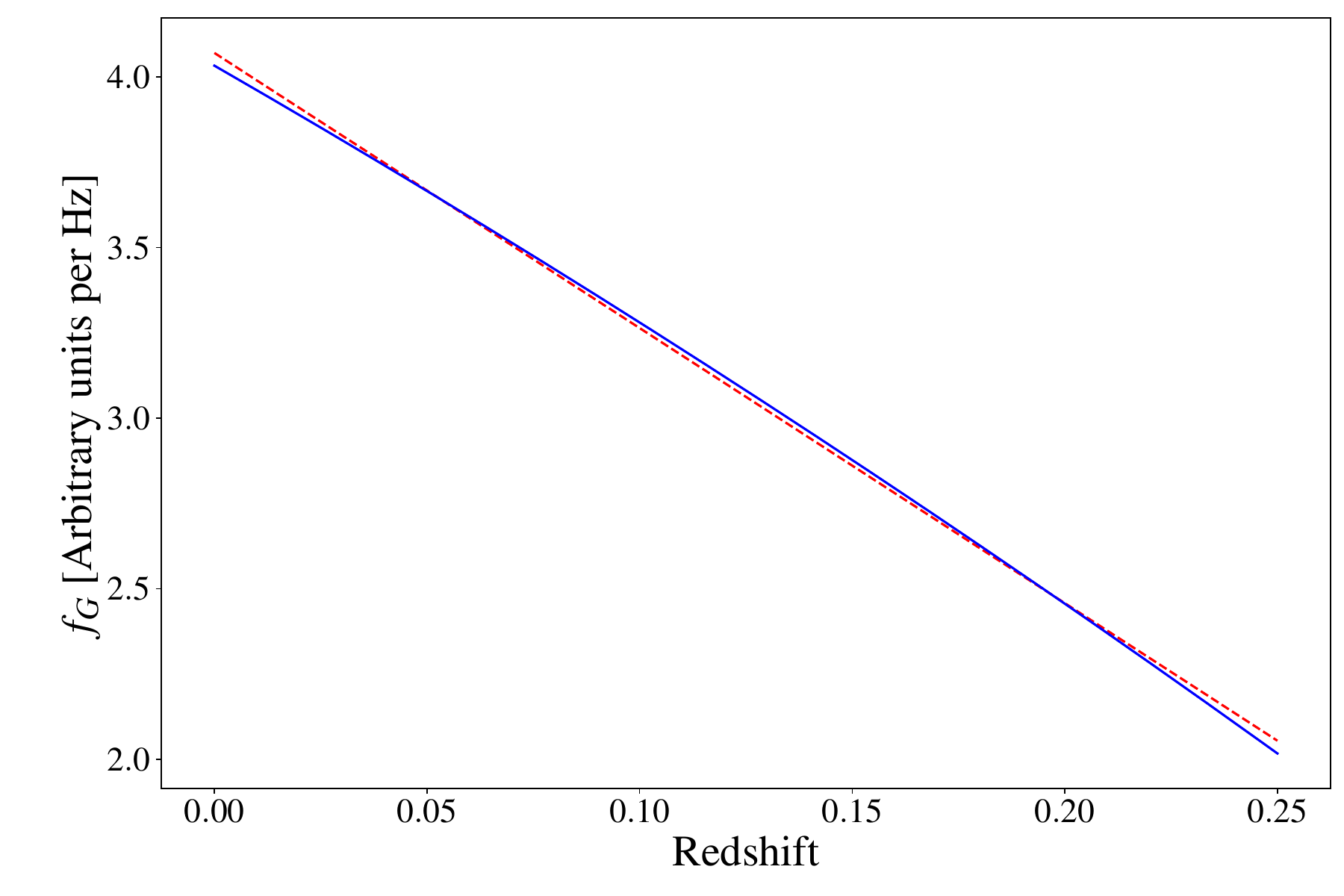}
	\caption{\label{fig:toy}Flux in the $G$ passband as a function of redshift (blue solid line), assuming $f_U=1$ and $f_R=2$ in arbitrary units per Hz. The red dashed line shows a linear fit to the relation.}
\end{figure}

We build here a toy model to allow us to estimate {the amplitude of} {the effect of passband variations} on the \phz\ using a simplistic SED consisting of a step function (when expressed per {wavelength}) at the wavelength of the Balmer break, which we set here to be exactly 400\,nm. We consider a system of three top-hat transmission curves $UGR$, with $U(\lambda)=1$ if 300\,nm\,$<\!\lambda\!<$\,400\,nm, $G(\lambda)=1$ if 400\,nm\,$<\!\!\lambda\!\!<$\,500\,nm, and $R(\lambda)=1$ if 500\,nm\,$<\!\!\lambda\!\!<$\,600\,nm, all transmission curves being 0 outside of these ranges. From Eq.~\eqref{eq:flux_ab}, we find that the fluxes $f_U$ and $f_R$ are constant if we consider only redshifts $z<0.25$, since the Balmer break remains within the $G$ band. As the redshifted Balmer breaks moves across the $G$ passband, $f_G$ changes as a function of $z$ according to Eq.~\eqref{eq:flux_ab} (see Fig.~\ref{fig:toy}), so that there is a direct relationship between $f_G$ and the redshift. Because of the $\lambda$ term in the numerator of Eq.~(\ref{eq:flux_ab}), the relationship is not exactly linear.

\begin{figure}
	\includegraphics[width=8.8cm]{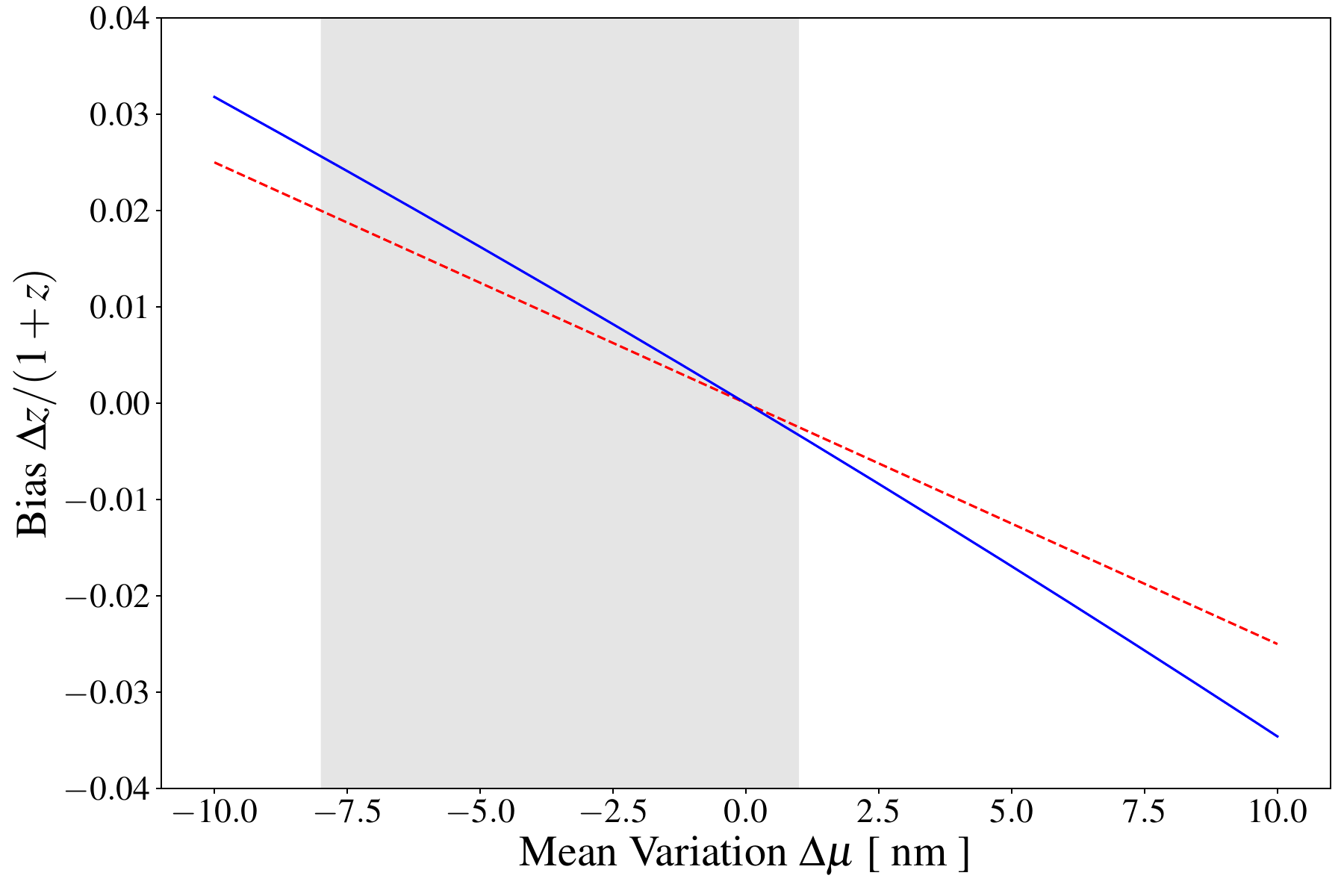}
	\caption{\label{fig:toy_mean}Bias $\Delta z$ resulting from a shift of $\Delta\mu$ in the $G$ passband for $z=0.15$ (solid blue line). The dashed red line is the `theoretical' bias $\Delta\mu/400$\,nm, where $\Delta\mu$ is the error on the location of the Balmer break. The grey area shows the domain of variations of the mean wavelengths of the measured MegaCam $r$ passbands (from 0 to 8\,nm).}
\end{figure}

We consider a varying $G$ passband where the only possible variation is a shift in the {mean} wavelength of the passband by an amount of $\delta$, i.e.\ $G(\lambda)=1$ if 400\,nm\,$<\!\lambda-\delta\!<$\,500\,nm, and  $G(\lambda)\!=\!0$ otherwise. Obviously, the mean $\mu$ of $G$ is shifted by the same amount $\Delta\mu=\delta$. As discussed in Sect.~\ref{sec:measurement}, typical values of $\delta$ can be of the order of a few nanometres. Figure~\ref{fig:toy_mean} shows the bias that results from a shift of $\Delta\mu$ of the passband for a true redshift $z=0.15$.

From Fig.~\ref{fig:toy_mean} we found that the bias $\Delta z$ is almost a linear function of $\mu$, {with $\Delta z=-0.0033\Delta\mu/1$\,nm}. This is close, but not identical, to the expected bias if one is able to locate the Balmer break with an error of $\Delta\mu$. In such a case, we would have: $\Delta z=\Delta\mu/400$\,nm${=-0.0025\Delta\mu/1}$\,nm. {The larger slope is due to the stronger weight of long wavelengths in Eq.~\eqref{eq:flux_ab}}. This very simplistic analysis shows nevertheless that a shift of 1\,nm leads to a bias that is of similar amplitude to \Euclid's requirement on the bias, $\Delta z/(1+z)=0.002$, which shows that changes in passband variations need to be taken into account in the computation of \phz s. {In Fig.~6, we also showed the range of $\Delta\mu$ observed in Sect.\ref{sec:measurement}. We found that, because of the tendency of off-axis transmissions to move towards the blue, \phz s are expected to be biased positively if the central passbands are used.}

\subsubsection{Consequences for \phdz\ algorithms}
Passband variations add a major complexity in the process, because potentially each source is observed with a (slightly) different set of passbands, which means that each source lives in a different colour space (which we assume is known through the measurements of its actual passband). Therefore each source requires a different mapping from colour space to redshift. The two main approaches to \phdz\ determination face significant, but distinct, issues when dealing with multiple colour spaces.
\paragraph{Template-fitting}
\label{sec:effect}
The TF approach to determine \phz s \citep[e.g.,][\!\!\!\!; Paltani et al. in preparation]{Arnouts1999,Bolzonella2000} involves the knowledge of galaxy SEDs, the so-called templates, which are assumed to be known at all relevant rest-frame wavelengths. The source fluxes are compared with reference fluxes that are computed by integrating the templates through the passbands. Hence, it is straightforward to compute the reference fluxes in the colour space of the source of interest. However, the calculation of the reference fluxes implies integrations through passbands and becomes computationally expensive for any reasonably large survey, and impossible for catalogues of billions of sources such as the future \Euclid Wide survey. In the case where all sources occupy the same colour space, this is solved by first calculating a single grid of fluxes for all models at all redshifts. {If each passband has $n$ variations, the model grid becomes $n$ times larger, which could become quite cumbersome}.

\paragraph{Machine-learning approaches}
\label{sec:ml}
ML is a vast class of algorithms that share the basic principle to infer the desired relation (in this case between the source's fluxes and the redshift) in a purely data-driven manner. ML {approaches} almost always involve a training phase, where the algorithms build up internally (learn) the colour-redshift relation. After the training phase, \phz s can be very efficiently computed through this newly determined relation. However, the training phase of ML algorithms can be a quite computationally intensive process, depending on the algorithm. This is not an issue in the case where all the sources and reference objects occupy the same colour space, since it needs to be performed only once. However, in the presence of passband variations, all sources can in principle occupy distinct colour spaces, so that the training phase would need to be performed many times, which may become computationally difficult. Another difficulty lies in finding enough {training objects} in each colour space, so that the colour-redshift relation can be accurately learned. As a matter of fact, finding a reference sample that covers entirely a single colour space of galaxies is already extremely difficult \citep{Masters2015}{.}

\section{Implementation in template-fitting algorithms}

{The TF algorithm is based on the assumption that the SEDs of the real objects are drawn from a known set of SEDs, so that we are able to compute the predicted model colours in any colour space. However, associating to every source the full passband information, including the variations specific to this source, is quite demanding in terms of data management. Furthermore, this results in a much larger model grid size. We thus propose here a simplified correction that only takes into account the shifts in the {mean} wavelengths of the passbands, and we then validate our approach with simulations.}




\subsection{Template-fitting likelihood}
{TF algorithms compare} the source fluxes with those obtained from simulated objects with known parameters $\alpha$. These parameters are typically {the set of reference SEDs used to match the observed SED of the source, the redshift of the source, the internal reddening law (e.g., \citealt{Prevot1984}, \citealt{Calzetti2000}, etc.), and the value of internal reddening $E_{B-V}^\mathrm{int}$}. The likelihood of the match as a function of $\alpha$ is given by $\exp{\left(-\chi^2_\alpha/2\right)}$, with
\begin{equation}
\label{eq:chi2}
\chi^2_\alpha=\sum_{T\in {\cal T}} \frac{\left(f^\mathrm{s}_T-ar^\alpha_T\right)^2}{{\sigma^\mathrm{s}_T}^2},
\end{equation} 
where the sum runs over all passbands ${T\in {\cal T}}$, $f^\mathrm{s}_T$ is the source flux through passband $T$, $r^\alpha_T$ is the reference flux of the simulated object with parameters $\alpha$ obtained from Eq.~\eqref{eq:flux_ab}, and $\sigma^\mathrm{s}_T$ are the uncertainties of the source fluxes in passband $T$. Finally, $a$ is a scale factor that is left free to minimise $\chi^2_\alpha$ in Eq.~\eqref{eq:chi2}; alternatively, $a$ can be included in $\alpha$, so that it can be marginalised upon or its posterior can be obtained (this is especially useful for the determination of physical parameters; see, e.g., \texttt{Phosphoros}\footnote{Available using \texttt{Anaconda}; see\\ \url{https://anaconda.org/astrorama/phosphoros}}; Paltani et al., in preparation).

\subsection{Correction factor for passband variations}
\label{sec:corr}
The likelihood in Eq.~(\ref{eq:chi2}) does not take into account the possibility of filter variations. If the source flux is measured through a variation $T'$ of passband $T$, and the simulated flux with parameters $\alpha$ is computed using passband $T$, we introduce a correction factor $C^\alpha_{T\rightarrow T'}$ to be applied to $r^\alpha_{T}$ in order to obtain $r^\alpha_{T'}$ to be used in Eq.~(\ref{eq:chi2}) as
\begin{equation}
\label{eq:fluxcorr}
r^\alpha_{T'}=C^\alpha_{T\rightarrow T'}\,r^\alpha_{T}\;,
\end{equation}
so that we get a new equation for $\chi^2_\alpha$:
\begin{equation}
\label{eq:chi2_corr}
\chi^2_\alpha=\sum_{T\in {\cal T}} \frac{\left(f^\mathrm{s}_{T'}-a\,C^\alpha_{T\rightarrow T'}\,r^\alpha_{T}\right)^2}{{\sigma^\mathrm{s}_{T'}}^2}\;.
\end{equation}

Since we know the SED $s^\alpha(\lambda)$ and the passbands $T$ and $T'$, we can use Eq.~(\ref{eq:flux_ab}) to determine the correction factor $C^\alpha_{T\rightarrow T'}$ :
\begin{equation}
\label{eq:corr_factor}
C^\alpha_{T\rightarrow T'}=\frac{r^\alpha_{T'}}{r^\alpha_{T}}= \frac{\int_0^\infty s^\alpha(\lambda)\, T'(\lambda) \,\lambda \,\mathrm{d}\lambda}{\int_0^\infty s^\alpha(\lambda)\, T(\lambda) \,\lambda \,\mathrm{d}\lambda}~\frac{\int_0^\infty T(\lambda)\,\frac{\dd\lambda}{\lambda}}{\int_0^\infty T'(\lambda)\,\frac{\dd\lambda}{\lambda}}\;.
\end{equation}

The assumption that the main parameter affecting the bias is the {mean} wavelength of the passband allows us to propose a very important simplification. With $\Delta\lambda$ being the difference of {mean} wavelength between $T$ and $T'$,
\begin{equation}
	\label{eq:dlambda}
	\Delta\lambda=\int_\lambda \left[T'(\lambda)-T(\lambda)\right]\,\lambda\,\dd\lambda\;,
\end{equation}  
the correction factor can be expressed as a function of $\Delta\lambda$, that is, $C^\alpha_{T\rightarrow T'}=C^\alpha_T(\Delta\lambda)$. We can thus compute $C^\alpha_{T}(\Delta\lambda)$ for different wavelength shifts of the passband $T$ using Eq.~(\ref{eq:corr_factor}). {We note that $\Delta\lambda$ is simply the difference between the mean wavelengths of the two passbands, and, crucially, does not depend on the SED}. In the case of multiple exposures with different variations of the passband, the resulting $\Delta\lambda$ is defined as the exposure time-weighted average of the individual $\Delta\lambda_i$ of the different exposures. We point out that this approach is very similar to the correction of the Galactic extinction using the full knowledge of the SED as developed by \citet{Galametz2017} (see Appendix \ref{sec:ebv} for more details), which is implemented in \texttt{Phosphoros} (Paltani. et al., in preparation).

Appendix~\ref{sec:corr_factor} describes in detail how $C^\alpha_{T}(\Delta\lambda)$ can be approximated with an analytical function of $\Delta\lambda$. We found that we get excellent approximations of $C^\alpha_{T}(\Delta\lambda)$ for all sets of parameters $\alpha$, all redshifts, and all four CWW templates using a {second-order} polynomial with the constant term fixed to 1:
\begin{equation}
\label{eq:deg2fit_main}
C^\alpha_{T}(\Delta\lambda)\approx a^\alpha_{T}\Delta\lambda^2+b^\alpha_{T}\Delta\lambda+1\;.
\end{equation}
{Since $a^\alpha_{T}$ and $b^\alpha_{T}$ depend only on the passband $T$ and on the set of parameters $\alpha$, we can precompute grids of these parameters.}


\subsection{Simulations}
{In order to validate our approach,} we performed idealistic, {noiseless} simulations to estimate the bias resulting from passband variations. We used the four CWW templates \citep[Elliptical, Sab, Sbc, Irregular;][]{Coleman1980} in order to estimate the bias over a range of galaxy types. We simulated MegaCam $ugriz$ and \Euclid \YE\JE\HE\ photometry of objects modelled with the CWW templates ignoring internal reddening and photometric uncertainty over the redshift range 0--3. We chose MegaCam because of its rather large passband variations; in addition, MegaCam $u$ and $r$ passbands {will} be used for the northern part of the \Euclid wide survey \citep{Scaramella2022}. Objects were simulated in each band with randomly chosen instances of its ten possible MegaCam variants (see Sect.~\ref{sec:megacam}). For the \Euclid NISP passbands, we {created} ten arbitrary passbands by shifting the nominal passband by {0} to 8\,nm. We then determined the redshift using TF using only the central passbands, ignoring passband variations. In absence of passband variations, this setup would produce perfect photometric redshifts, without any uncertainty, nor bias, so that any uncertainty or bias is entirely due to the mismatch between the passbands used to determine the source and reference fluxes, respectively. We performed three different tests of increasing complexity: {firstly}, only the $r$ passband can vary; secondly all passbands can vary {independently}; and finally fluxes are measured using a stack of four exposures, each of them having random {sets of} passband variations.

\begin{figure*}
	\includegraphics[width=6cm]{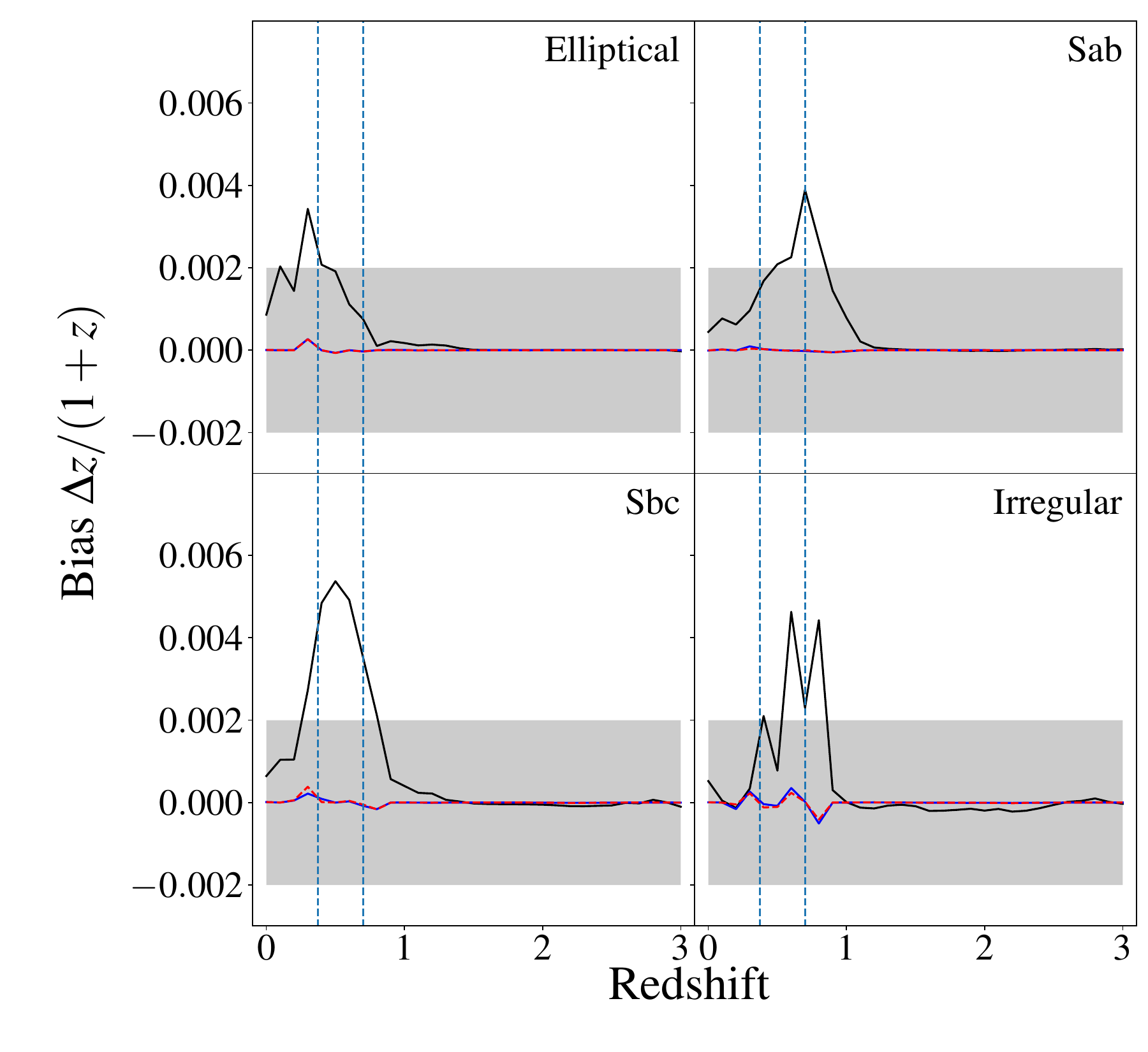}%
	\includegraphics[width=6cm]{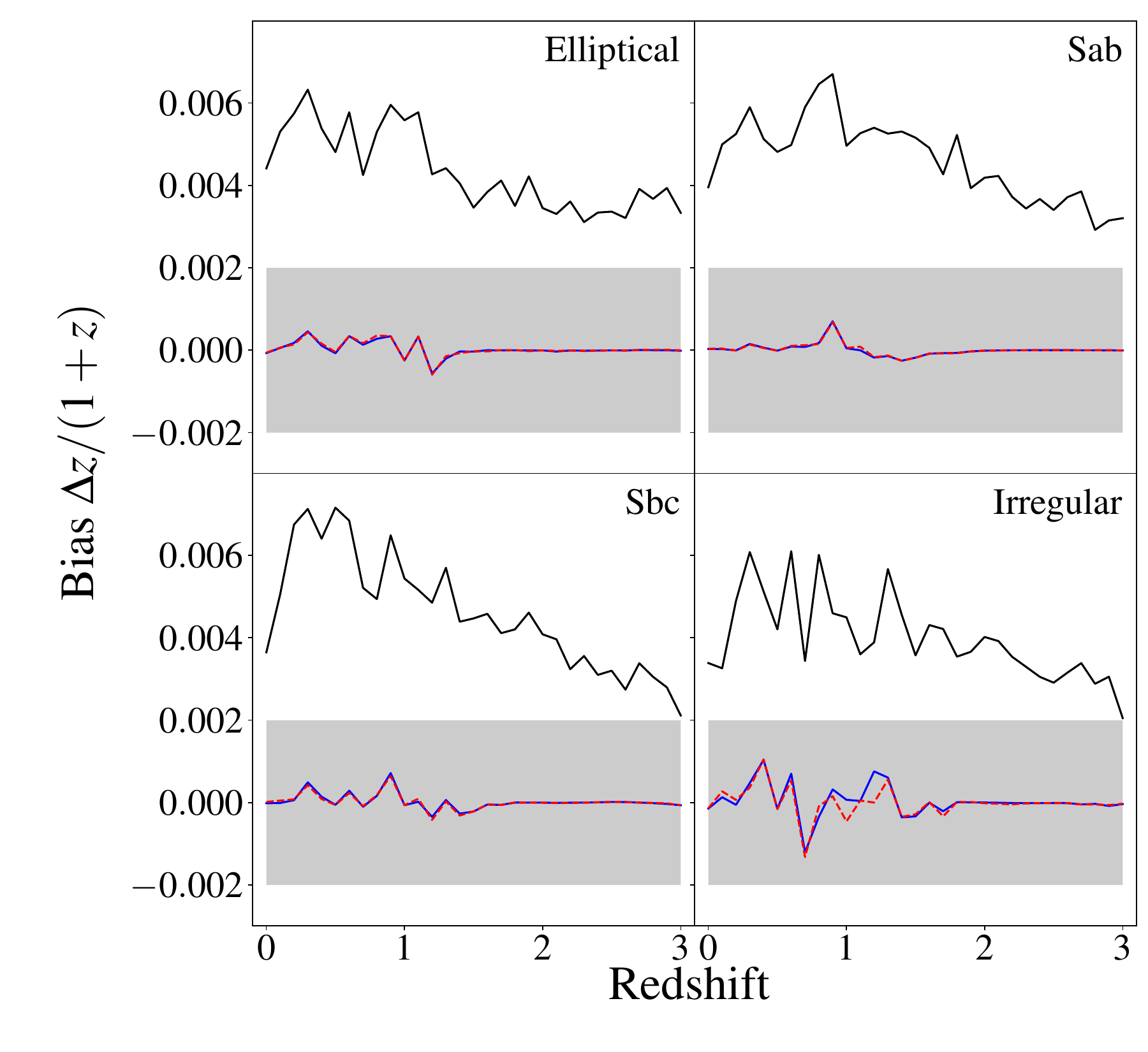}%
	\includegraphics[width=6cm]{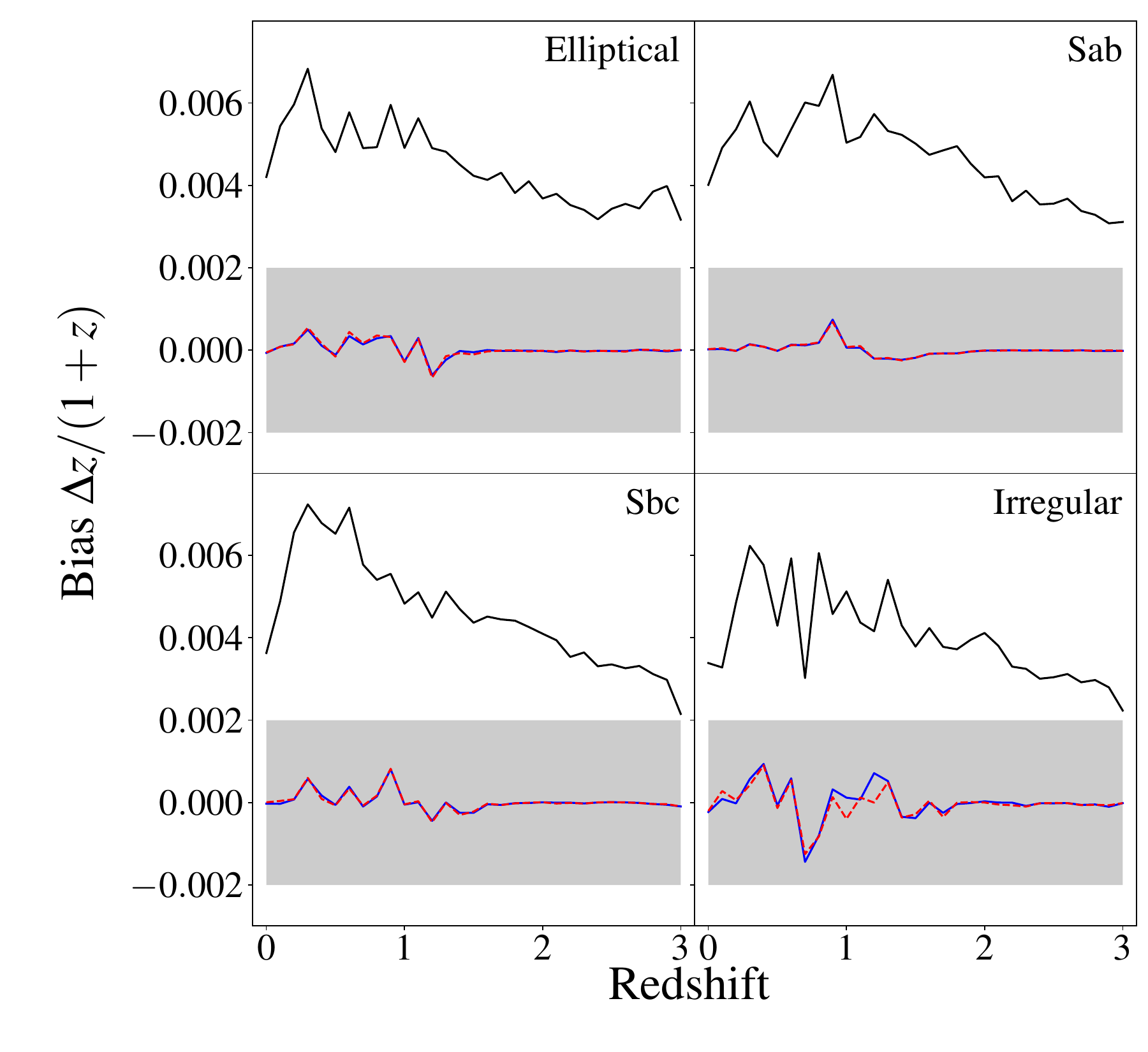}
	\caption{\label{fig:sim-bias}Bias in the \phz s {due to passband variations} as a function of redshift in the MegaCam $ugriz$ + \Euclid \YE\JE\HE\ configuration. Each plot shows the bias for the four CWW templates, as indicated. \emph{Left}: only the $r$ passband is variable; \emph{centre}: all passbands are variable; \emph{right}: all passbands are variable, and four exposures have been stacked. {In the left plot, the vertical dashed lines indicate the redshifts where the Balmer break enters and exits the $r$ passband}. The different {curves} show the bias when the passband variations are ignored (black line), or when {the full central passbands corrected for the shifts in {mean} wavelengths (blue line) are used}. The dashed red lines show the bias {when the second-order polynomial correction on the flux} presented in Sect.~\ref{sec:corr} is used. We point out that the blue and red lines are very close, so that the blue line is {often} barely visible.}
\end{figure*}

Figure \ref{fig:sim-bias} shows the resulting bias in the three configurations. When the central passbands are used {to determine the \phz s}, some bias is clearly present at a level that is of the same order as the accuracy requirement for \Euclid in all three configurations. When only the $r$ passband is randomised, {the bias is completely concentrated in a specific redshift range, which matches quite well that where the Balmer break, which is present to different extents in all four CWW templates, falls into the $r$ passband. When all passbands vary, {the bias is present at all redshifts and reaches about $0.007(1+z)$ in the worst case. Considering an average passband shift of 5\,nm, in our toy model the bias reaches about $0.011(1+z)$ at $z=0.15$, which is about 50\% larger than the bias we found in the simulations}. This difference was expected because the features in the CWW templates are not as sharp as the step function we used in our toy model; the fact that we are at redshift $\sim\!0.5$ instead of 0.15 further smooths the transition. When multiple exposures were stacked, the bias was practically identical to that in the single-exposure case because using four exposures makes the effective transmission less variable, without changing its average.} 

\begin{figure*}
	\includegraphics[width=6cm]{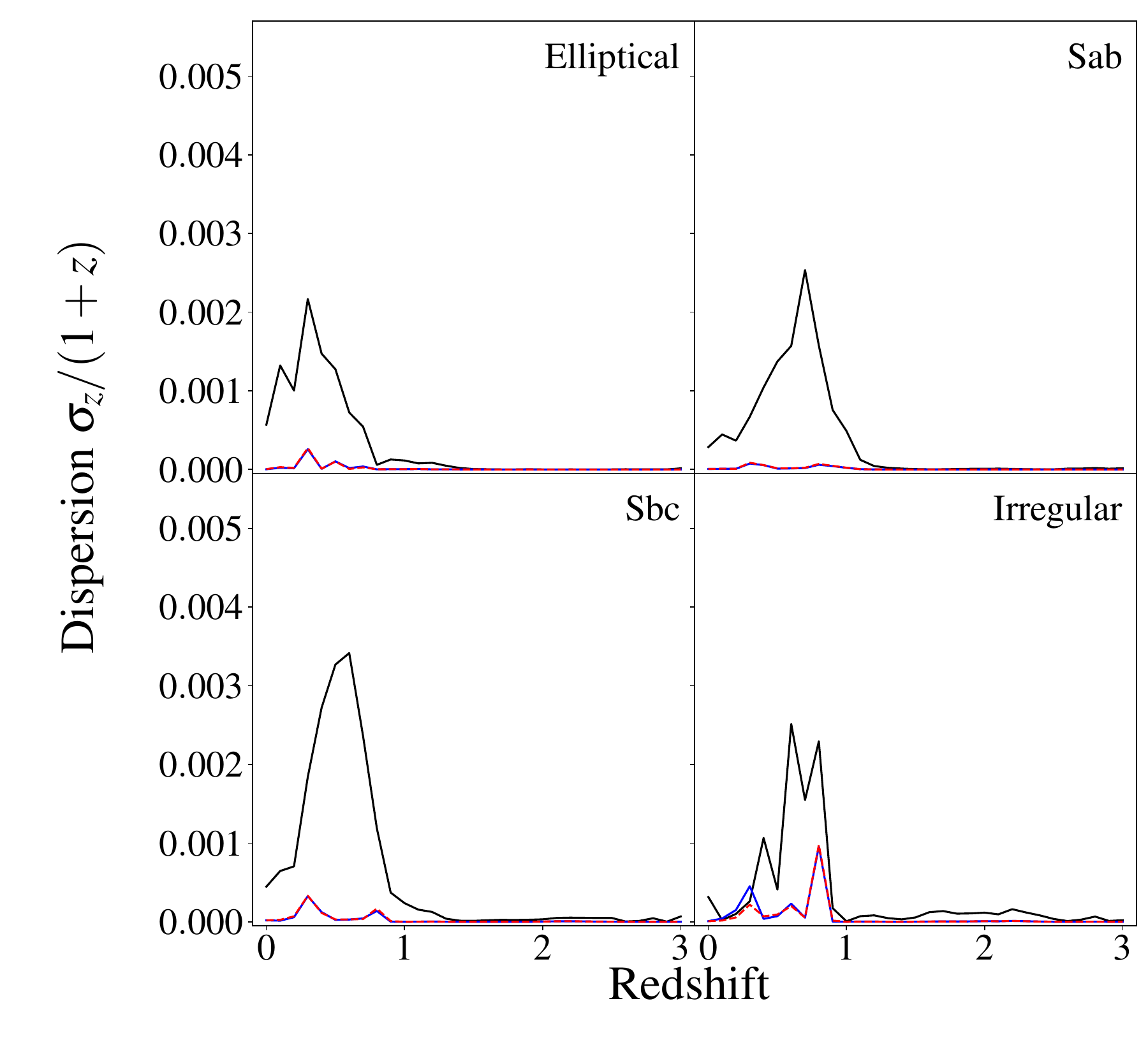}%
	\includegraphics[width=6cm]{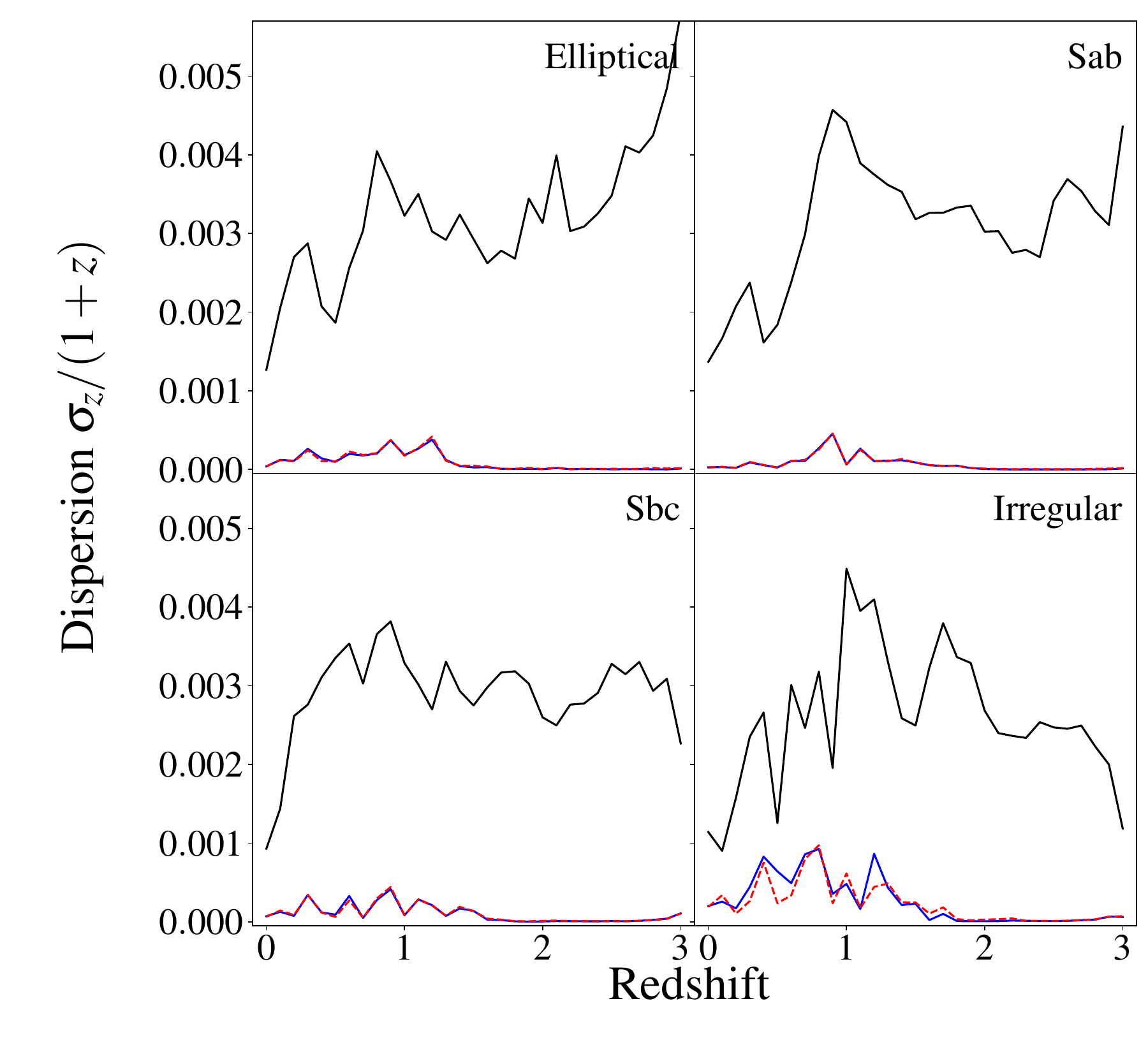}%
	\includegraphics[width=6cm]{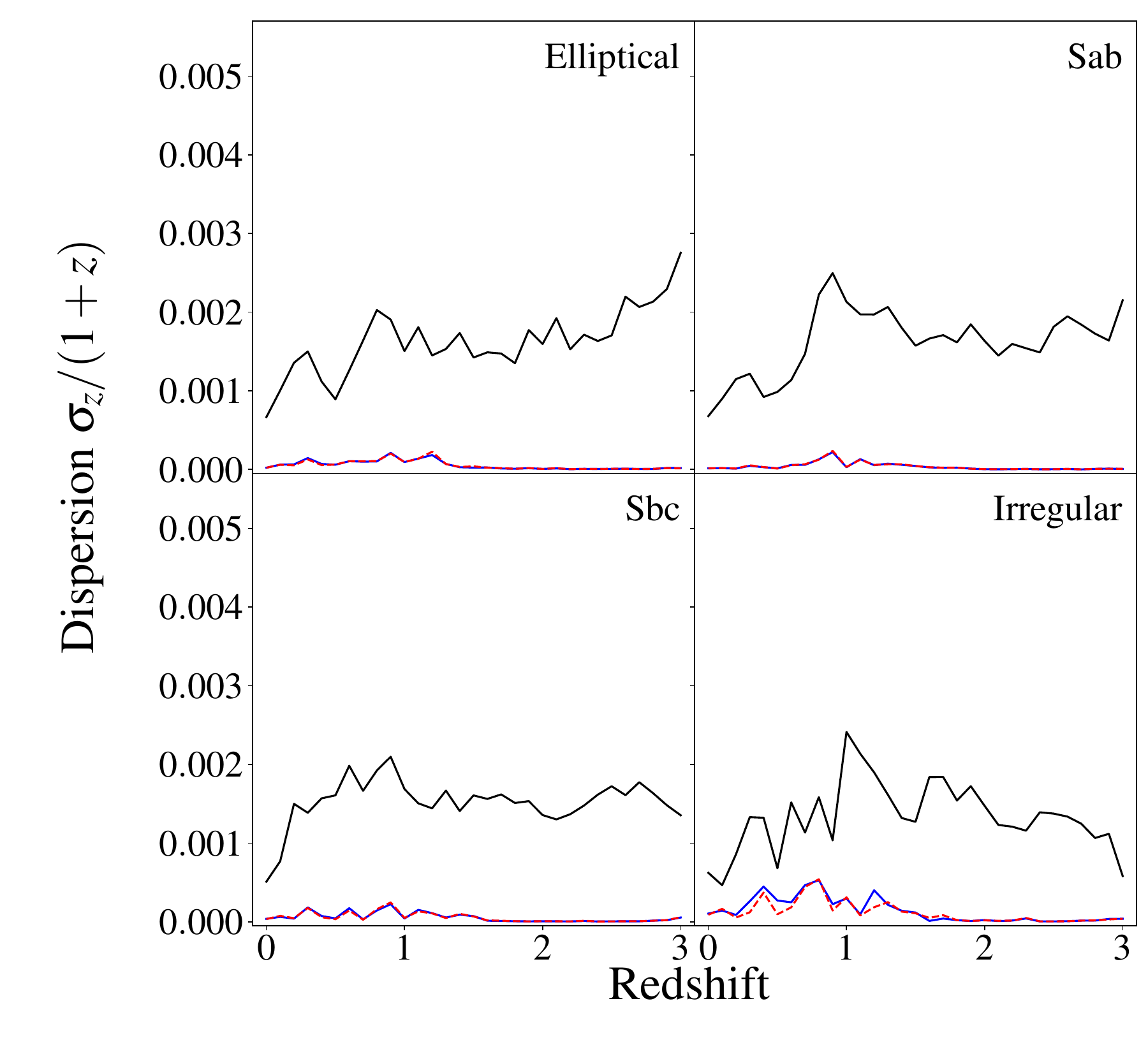}
	\caption{\label{fig:sim-sigma}Dispersion in the \phz s {due to passband variations} as a function of redshift in the MegaCam $ugriz$ + \Euclid \YE\JE\HE\ configuration. Each plot shows the dispersion for the four CWW templates, as indicated. \emph{Left}: only the $r$ passband is variable; \emph{centre}: all passbands are variable; \emph{right}: all passbands are variable, and four exposures have been stacked. The different {curves} show the bias when the passband variations are ignored (black line), or when {the full central passbands corrected for the shifts in {mean} wavelengths (blue line) are used}. The dashed red lines show the bias {when the second-order polynomial correction on the flux} presented in Sect.~\ref{sec:corr} is used.  We point out that the blue and red lines are very close, so that the blue line is barely visible.}
\end{figure*}

Figure \ref{fig:sim-sigma} shows plots similar to Fig.~\ref{fig:sim-bias}, but for the dispersion. We see again that using wrong passbands has an effect on the quality of the \phdz\ predictions. This effect, which reaches at most $0.005(1+z)$ is however quite small compared to the \Euclid requirement on the dispersion [$\sigma_z=0.05(1+z)$] in all configurations. We note that, in the case of four exposures, the dispersion is a factor 2 lower, as expected from the averaging of four exposures. 


{When we applied the $C^\alpha_T(\Delta\lambda)$ correction factors, we found that the bias was significantly reduced, such that it always remained within the requirements (see Fig.~\ref{fig:sim-bias}). It is even in general below $0.0005(1+z)$, except in the case of the `Irregular' SED, where a peak at about $0.0015(1+z)$ remains. This is probably {due} to the presence of sharp features, such as strong emission lines, in this SED. In Fig.~\ref{fig:sim-sigma} we see that using only the shifts in mean wavelengths was also able to reduce the dispersion to a large extent, except in the case of the `Irregular' SED, where some residual dispersion remains. This is again probably an effect of the presence of sharp features, such as strong emission lines, in this SED.}  As a conclusion, for realistically varying passbands, the knowledge of the full passbands for each objects is not necessary; it is sufficient to know the {mean} wavelengths of the passbands, which is a quantity that is much easier to handle by \phdz\ algorithms, and in particular TF. The bias and dispersion found when using the second-order polynomial approximation of $C^\alpha_T(\Delta\lambda)$ were extremely close to those involving the full $C^\alpha_T(\Delta\lambda)$, demonstrating that the second-order polynomial approximation provides a very good representation of the correction factors.

\section{Application to real data}
\label{sec:appli}
We verified the capability of the method developed here to reduce the bias in the \phdz\ determination by applying it to real data. We used the seventh (final) data release of the Legacy Survey performed at the Canada-France-Hawaii Telescope (CFHTLS\footnote{\url{http://terapix.calet.org/terapix.iap.fr/rubrique5c64.html?id_rubrique=268}}). We used only the W1 wide-field, which we matched with the VIPERS Public Data Release 2 \citep[VIPERS-PDR2;][]{Scodeggio2018} spectroscopic-redshift catalogue obtained with the VIMOS multi-object spectrograph on the ESO-VLT; {VIPERS is colour-selected to include mostly sources at redshifts $0.5<z<1.2$} \citep{Garilli2014}. The CFHTLS catalogue contains the {MegaCam versions of the usual} $ugriz$ passbands, with some objects being observed with a different $i$ passband  denoted $y$\footnote{This $y$ band mostly {overlaps} the $i$ band, and should not be confused with the usual $y$ or \Euclid \YE\ bands, which lie at the limit between optical and near-infrared.}. Since we were mostly interested in the bias, we selected sources brighter {than} $r=21.5$, in order to remove as much as possible the statistical uncertainties from the \phdz\ determinations. From the VIPERS-PDR2 catalogue, we only kept very secure objects with flags either 3.5 or 4.5, excluding stars (identified with a spectroscopic redshift of 0). The match between the two catalogues resulted in 4915 objects. {With the cut at bright magnitudes, we found that most of the sources are found in the redshift range $0.4<z<0.7$.}

\begin{figure*}
	\includegraphics[width=18cm]{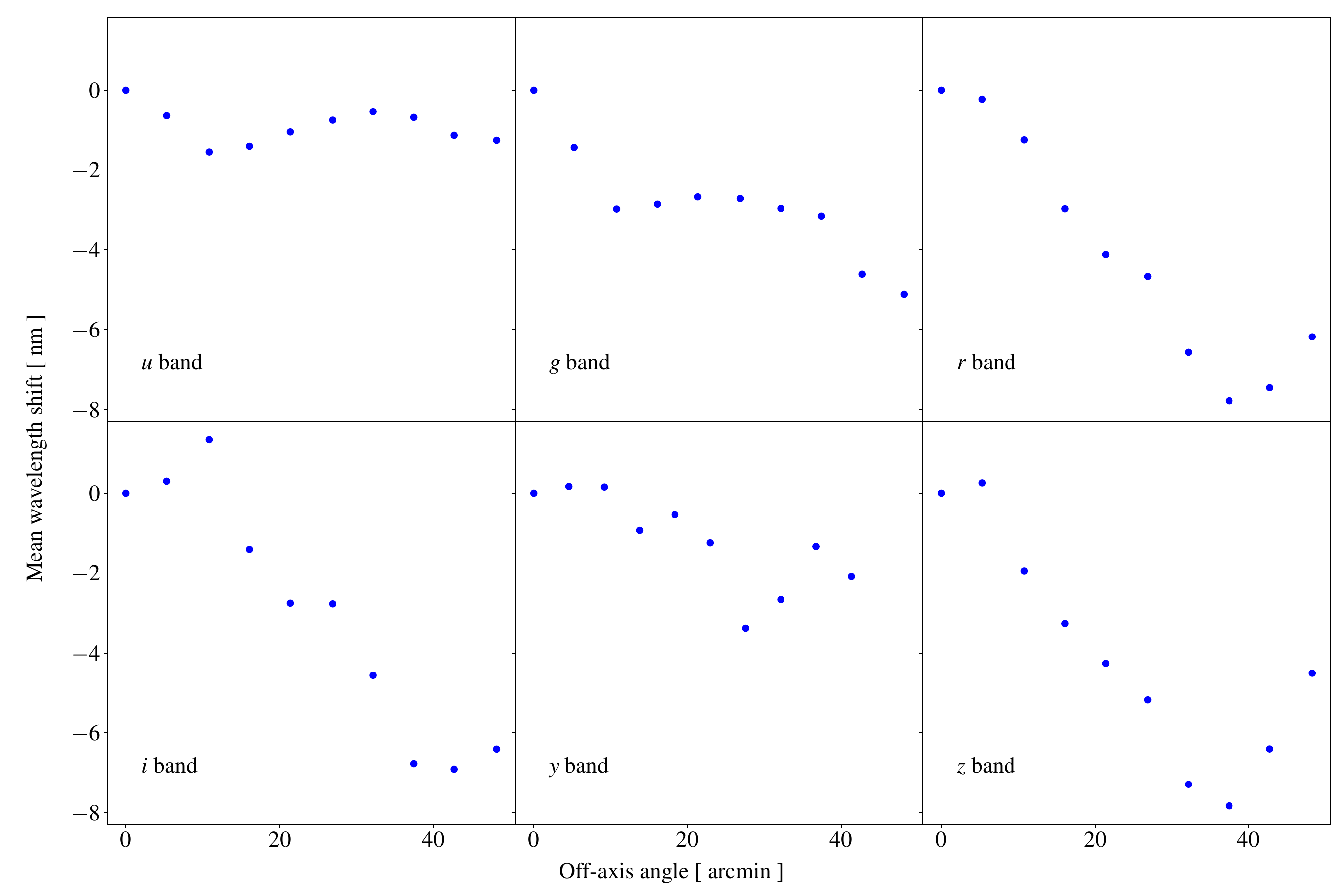}
	\caption{\label{fig:mega_offaxis}Mean wavelength shifts of the six MegaCam transmissions as a function of the off-axis angle compared to the on-axis transmissions. The $y$ transmission has been measured at slightly different off-axis distances from the other transmissions.}
\end{figure*}

\begin{figure}
	\includegraphics[width=8.8cm]{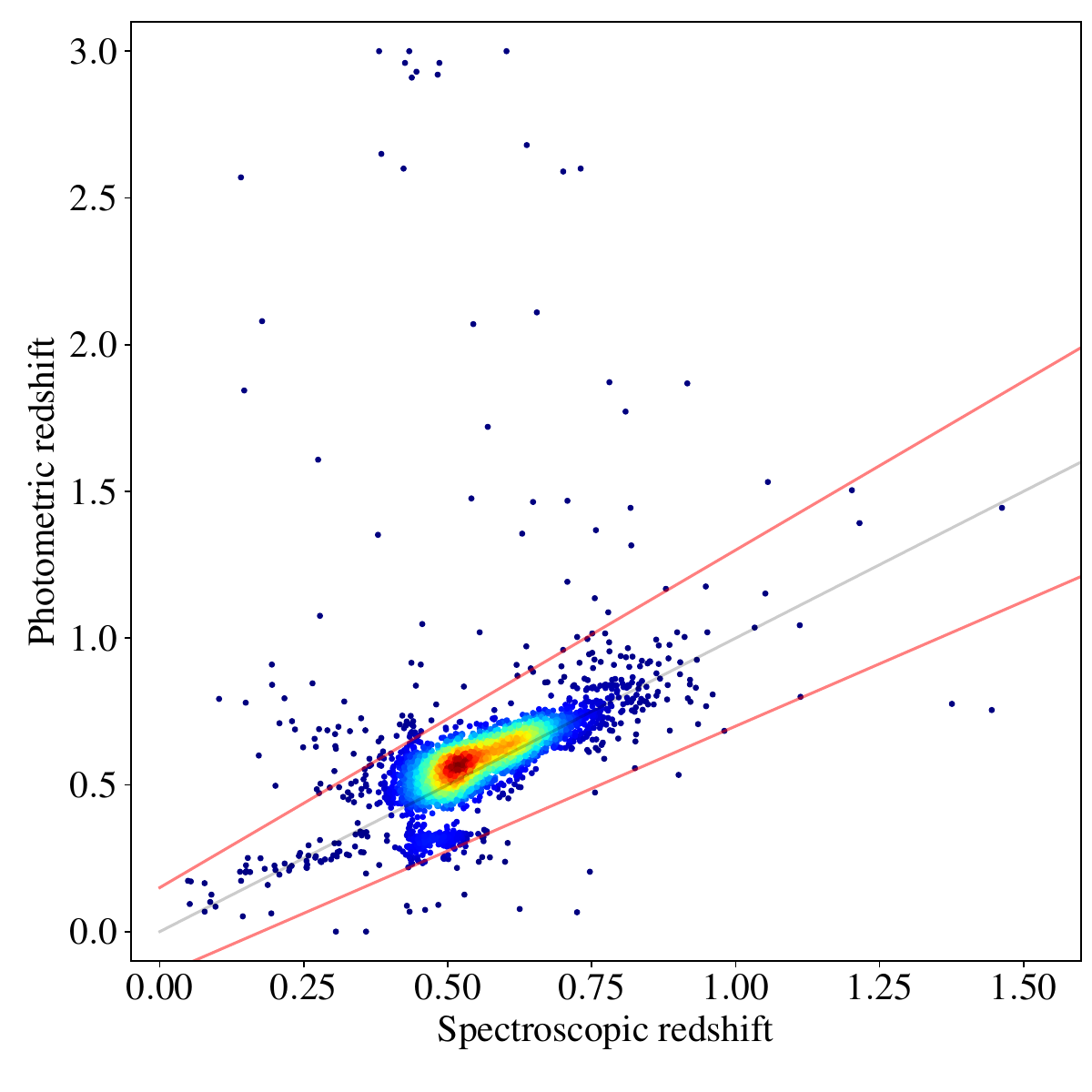}
	\caption{\label{fig:t007_phz}Photometric-redshift predictions for the W1 CFHTLS catalogue of bright galaxies with secure redshifts presented as a density plot in the \phdz--spectroscopic-redshift plane. The grey solid line is the 1:1 line. The red lines show the limits $0.85(1+z)$ and $1.15(1+z)$, respectively, traditionally separating good predictions from outliers.}
\end{figure}

{The calibrated data we were using do not contain any information regarding the atmospheric effects on the passband; consequently the only effect that we could take into account are the spatial variations of the passbands.} As discussed in Sect.~\ref{sec:megacam}, the transmission curves of the MegaCam instrument have been measured at ten different off-axis distances {\citep{Betoule2013}}. Using the pixel scale and pixel size of MegaCam, we could convert these physical distances into off-axis angles. {The pointing strategy of CFHTLS is such that a given object is observed with the same off-axis angles in all passbands, which maximises the effect of passband variations}. The shifts in the {mean} wavelengths of the passbands as a function of off-axis {angle} are shown in Fig.~\ref{fig:mega_offaxis}. The off-axis dependence is present in all passbands, but with quite different amplitudes. We note however that the relations are more complex than expected from purely geometric considerations, which means that the shift is spatially dependent in a more complex way that what we can model here.
We first computed the \phz s of the 4915 sources with a fully standard TF approach using \texttt{Phosphoros}. We used the 31 COSMOS templates used in \citet{Ilbert2009} and applied internal reddening up to $E_{B-V}^\mathrm{int}=0.5$ on the templates of spiral and starburst galaxies only using either the SMC extinction curve from \citet{Prevot1984}, or the \citet{Calzetti2000} extinction law for star-forming galaxies, with the addition of a bump at 2175\,\AA\ introduced by \citet{Massarotti2001}. Standard emission lines in \texttt{Phosphoros} have been added to the COSMOS templates based on the Kennicutt relation and the line flux ratios observed in sources in the SDSS-III/Baryonic Oscillation Spectroscopic Survey \citep[][\!\!\!; see Paltani et al., in preparation, for details]{Kennicutt1998,Thomas2013}. We do not apply any refinement in the algorithm, {such as a} luminosity prior, brightness prior, or zero-point correction, because we focussed on the determination of the bias, and not on the production of the best possible catalogue. Figure~\ref{fig:t007_phz} shows the overall quality of the \phz s. While the details of the performance are not important, we obtained very good predictions even with this limited analysis, with a dispersion (measured with the normalised median absolute deviation, NMAD) of 0.038 and an outlier fraction of 2.9\%. {The bias appears significant, with the region with the highest density of sources lying $\Delta z\sim 0.04$ above the 1:1 relation.}

\begin{figure*}
	\includegraphics[width=18cm]{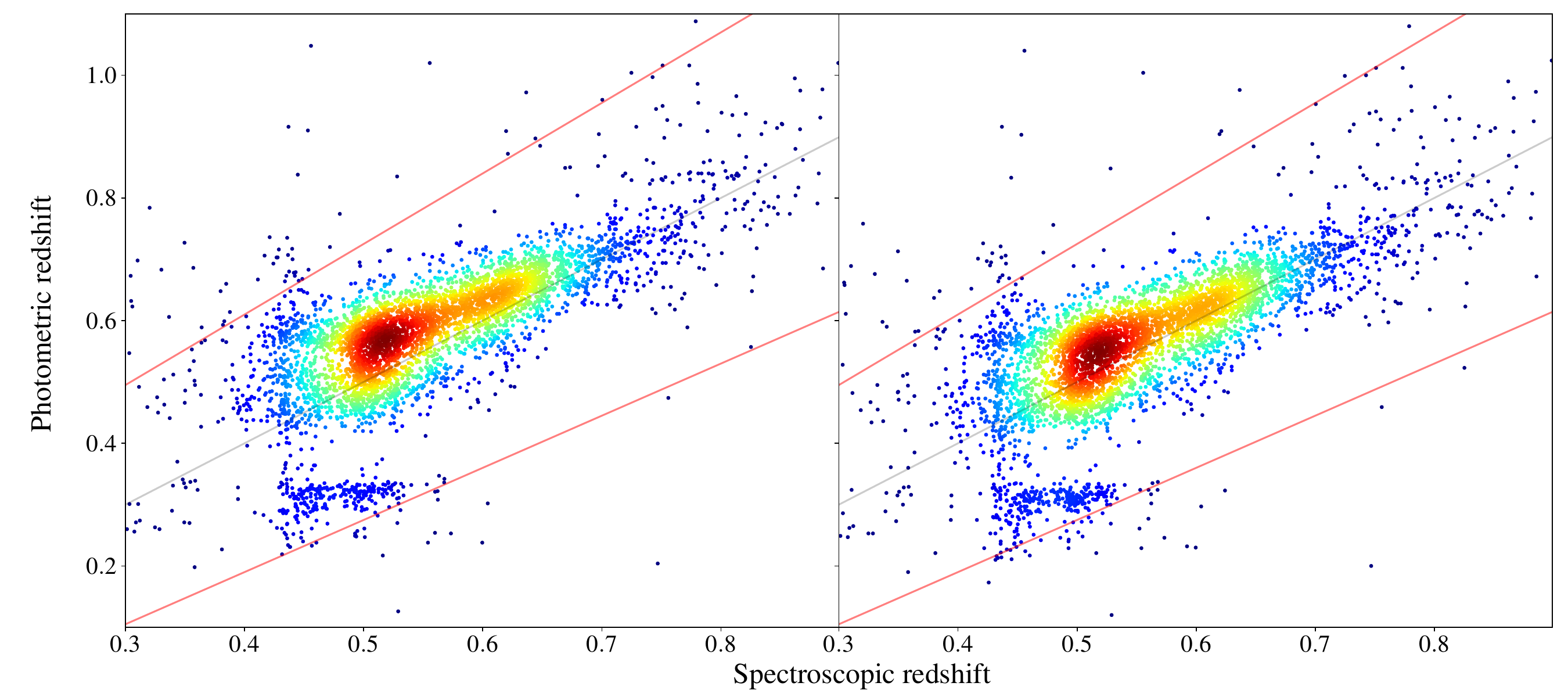}
	\caption{\label{fig:shift_noshift}Comparison of the \phdz\ predictions presented as density plots in the \phdz--spectroscopic-redshift plane without (left) and with (right) mean wavelength corrections. The grey solid line is the 1:1 line. The red lines show the limits $0.85(1+z)$ and $1.15(1+z)$, respectively.}
\end{figure*}

Using the off-axis angles of each source, we obtained wavelength shifts for each {passband} by interpolating the relations shown in Fig.~\ref{fig:mega_offaxis}. We used then \texttt{Phosphoros} in the exact same configuration as above, but this time taking into account the correction for wavelength shifts presented in Sect.~\ref{sec:corr}. We obtained a {normalised median absolute deviation NMAD} of 0.039 and an outlier fraction of 3.2\%. Both values are very {close to, but slightly worse than, those} obtained without the application of the correction for the mean wavelength shifts. The additional noise could result {from} the too simplistic assumption we made here that the wavelength shifts only depend on the off-axis angle. Figure~\ref{fig:shift_noshift} compares the densest part of the \phdz\ prediction plots with and without mean wavelength shifts. The bulk of the sources very clearly {moved} towards the 1:1 relationship when the shifts are applied, indicating a reduction in the bias, although some significant bias remains.

\begin{figure}
	\includegraphics[width=8cm]{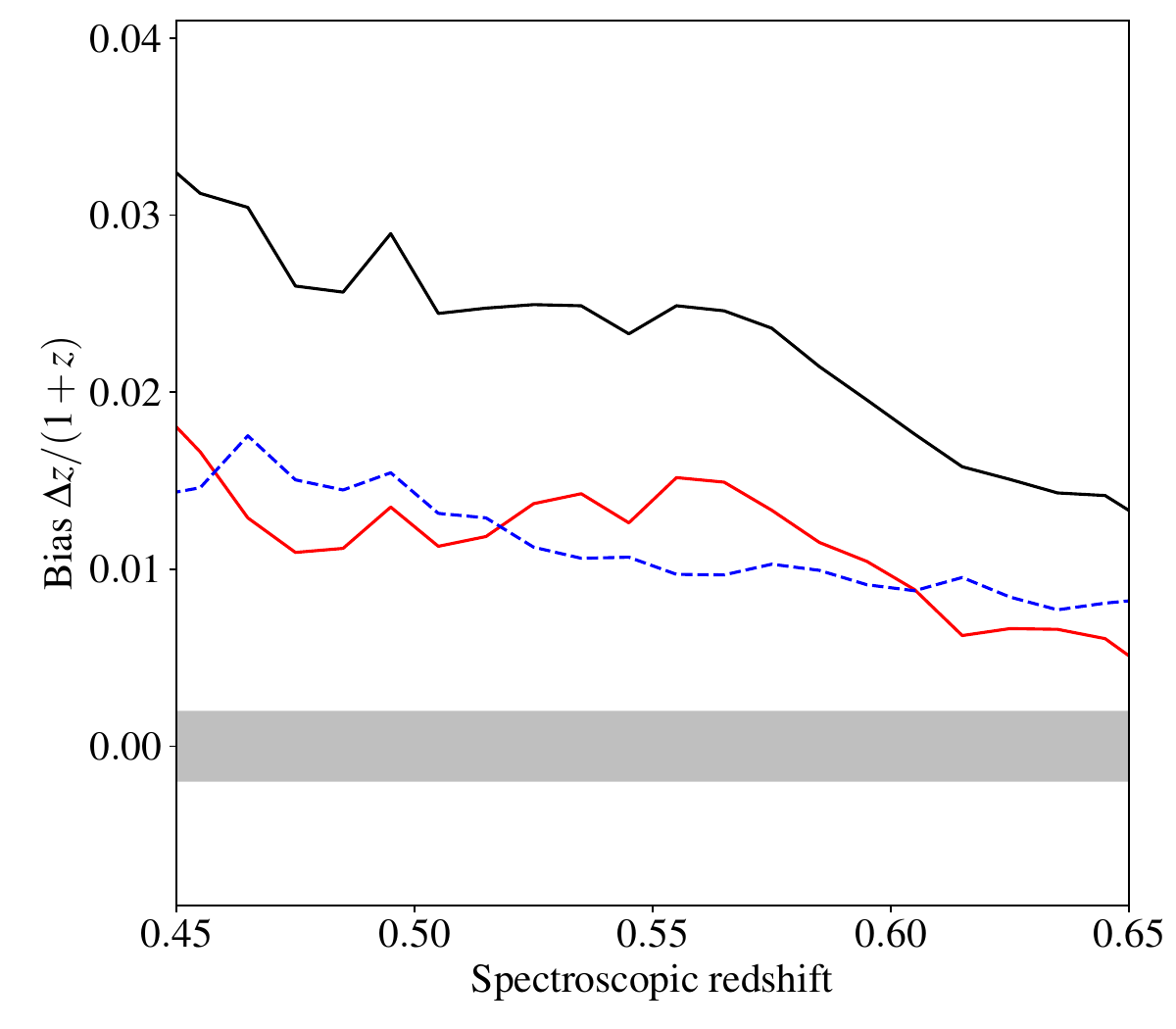}
	\caption{\label{fig:shift_bias}Median bias in \phdz\ predictions as a function of redshift for the original catalogue (black line) and when taking into account the correction in the mean passband wavelengths (red line). {The dashed blue line shows the reduction in the bias between the two solid curves.} The grey area indicates the \Euclid requirement. While it is clear that the passband correction reduces the bias significantly, there is still a need for a post-processing calibration step in order to meet the \Euclid requirements.}
\end{figure}

{We quantified more precisely the reduction in the bias by computing the median difference in redshift between the \phdz\ predictions and the true redshifts, in bins of redshift}. Figure~\ref{fig:shift_bias} compares the bias in the predictions from the original catalogue and from our computations involving the mean wavelength shifts over the redshift range $0.45<z<0.65$, where the bias can be reliably estimated. From Fig.~\ref{fig:shift_bias}, we found that the bias, which is positive everywhere, was reduced over this redshift interval by 0.018 in average. {As expected from the field-of-view dependence of the transmissions, passband variations induce a positive bias on the \phz s}. This reduction matches very well the expected bias due to passband variations for an average shift of the order of 5\,nm, which is the value expected in the case of the MegaCam passbands. However, at redshift $z= 0.48$, where the reduction reaches a maximum, the residual bias is about 0.016 and still exceeds the \Euclid requirement of $\Delta z/(1+z)<0.002$ by a factor about 8 \citep{Laureijs2011}. 

\section{Discussion}

\subsection{Origin of the residual bias in \phdz\ algorithms}
The method we proposed here is able to remove a significant fraction of the bias due to the variations of the passbands. {We designed the method to cope with spatial variations of the passbands, although in principle the same approach could be used for, for instance, passband variations due to atmospheric effects. In practice, this might be made difficult by the calibration process, which performs a colour-independent correction}. {In {Appendix}~\ref{sec:other} we describe other processes that can lead to (at least apparent) passband variations.} However, the residual bias is still larger than the requirements by a large factor. This is due to the fact that there are other sources of bias that are not due to the changes in the passbands. This could be due to other issues in the photometry. Any defect in the calibration process (for instance a wrong zero-point) will lead to biased determinations of the \phz. In the latter case, a zero-point correction \citep{Ilbert2006} can be introduced in the TF algorithms, and has been proven to be quite effective to reduce the bias. We did not use this correction here on purpose, since we wanted to focus on the effect of passband shifts. Other, in particular non-linear, issues in the photometry may not be easily removed. {One example is the imperfect PSF homogenisation that needs to be applied to the different photometric bands}. 

Another source of bias is the mismatch between the SED models and reality. For instance, SEDs are too few, or they do not match accurately enough the SEDs of true galaxies. To alleviate these issues, some \phdz\ codes include linear combination of templates \citep[\texttt{EAZY};][]{Brammer2008}, or SED templates can be adapted to match better with the observed colours \citep{Coupon2009}. We point out that the latter method may also be able to remove some bias inherent to the photometry. Other model errors that cannot be corrected by these methods may contribute to the bias, such as, for instance, imprecise estimate of the Galactic reddening, incorrect Galactic-reddening attenuation curve, inaccurate intrinsic reddening laws, or wrong assumptions on the IGM absorption.

Even if the models were correct, the priors could be incorrect, leading to biased estimates. In algorithms implementing Bayesian statistics the choice (or not) to apply a specific prior (e.g., luminosity function or similar, or the colour-space coverage of the SED models) leads to biases. Algorithms using maximum-likelihood are even more subject to bias, since the maximum-likelihood solution of an optimisation problem is biased when the error distribution is not symmetric. Even if one does not consciously use Bayesian statistics, some choices act as priors and affect the determination of the \phz s, such as, for example, the choice of a maximum value for the internal reddening or for a galaxy luminosity.

ML algorithms have in principle significant advantages over TF algorithms with respect to the bias, {as they can learn the defects in the photometry,} although we argued in Sect.~\ref{sec:effect} that passband shifts are difficult to take into account. {They also do not rely} on our imperfect knowledge of the Universe. However, the quality of the ML model depends a lot, and in a very complicated way, on how well the training sample matches the {target} data set. {In particular the requirement to get a spectroscopic redshift biases the training samples towards bright emission-line galaxies \citep[see ][for an in-depth discussion]{Hartley2020}. Thus ML algorithms are also affected by model imperfections and incorrect priors, which are hard to determine objectively}.

{Ultimately, a bias correction remains necessary, since not all sources of bias can be identified precisely enough to be corrected. Different approaches have been proposed to this end. In the case of \Euclid, a direct calibration in colour-space based on the approach developed in \citet{Masters2015} will be implemented. Such calibration at the level of the \Euclid requirements remains nevertheless extremely challenging \citep{Wright2020b}, so that any well-understood source of bias, such as the passband variation, should be removed beforehand as far as possible}.

\subsection{Comparison with the colour-term approach}

Using an approach similar to the colour-term calibration (see Sect.~\ref{sec:photom}), {\citet{Betoule2013} implemented a correction for the effect of passband variations based on a colour term. Using their notation (see their equation 6), the corrected magnitude $m_{|x_0}$ of a source with magnitude $m_{|x}$ is given by}
\begin{equation}
	\label{eq:betoule}
	m_{|x_0}\,\approx \,m_{|x}\,-\,\delta k(x)\,(c_{|x_0}-c^0(x)),
\end{equation}
{where $x_0$ refers to the centre of the field of view, $x$ is the location of the source in the field of view, $\delta k(x)$ is the position-dependent colour term, $c_{|x0}$ is the colour of the object, and ${c^0(x)}$ is the reference colour
for the affine transformation; we ignore here the flat-field and zero-point corrections, as they are not relevant for our discussion. \citet{Betoule2013} chose $c_{|x0}= (g -i)_{|x_0}$, which is a good temperature indicator for stars. In the limit of small passband variations, $\delta k(x)$ is small, which alleviates the need to obtain accurate colours; consequently, \citet{Betoule2013} ignore the effect of passband variations in the determination of $(g -i)_{|x_0}$.}

{We can compare the above equation with the correction in Eq.~(\ref{eq:fluxcorr}). Both approaches use an estimate of the SED of the source and some position-dependent factor to determine corrected flux or magnitude. A noticeable difference is that our correction is expressed as a function of a property, the mean wavelength shift, that is independent of the intrinsic properties of the object, {contrarily to the observed colour}. The {colour-term} correction is also based on a very crude approximation of the SED of the source based on a single colour, which may be unsuitable for the estimation of the SED for passbands that are far from the $g$ and $i$ passbands, such as the \Euclid \YE\JE\HE\ near-infrared passbands}. {By contrast, in our method the SED is determined over all passbands based on all the available photometry simultaneously using empirically or physically motivated templates. Eq.~\eqref{eq:betoule} also introduces an unwanted correlation with the $g$ and $i$ bands.}


One additional advantage of our approach is that {the SED is not needed during the photometric extraction and calibration stages}, which makes it easier to combine photometric catalogues from different surveys. In the case of the \Euclid survey of the southern hemisphere, the \Euclid near-infrared photometry will be {complemented} with the DES survey \citep[and with the Rubin LSST later on; see][]{Guy2022}. The northern sky will be even more complicated, with the optical survey consisting of observations from several telescopes \citep{Scaramella2022}: the CFHT \citep[Canada-France Imaging Survey;][]{Ibata2017} for the $u$ and $r$ bands; {Subaru with the Hyper Suprime-Cam \citep{Miyazaki2018} for the $g$ and $z$ band}; and Pan-STARRS \citep{Chambers2016} for the $i$ band. In order to constrain colour terms, the photometric-calibration approach would require the simultaneous analysis and calibration of all these photometric data, which can become impractical. In addition, it is now commonly accepted that the team {that} conducted a survey is able to provide the best calibration, so that a new scientific analysis rarely starts from the raw data, but rather from calibrated stacks, if not directly from the published photometric catalogues. Our approach can be applied in a straightforward way to any catalogue assembled from distinct catalogues while benefiting from the best possible calibration and passband variation correction. 

\subsection{Photometry corrected for passband variations}
\label{sec:corrphot}
The $C^\alpha_{T}(\Delta\lambda)$ correction factors can be computed only with TF algorithms, because these are the only \phdz\ algorithms for which the SEDs of the reference objects are known in general. ML algorithms require only a spectroscopic redshift. However, TF algorithms can be used to create corrected photometry {that} can later be used for any purpose, including the computation of \phz s using any other algorithms, simply by applying the $C^\alpha_{T}(\Delta\lambda)$ correction factor to the photometric flux and uncertainty through passband $T$ for the best-fit $\alpha$ parameters and the measured $\Delta\lambda$ wavelength shift. The possibility to provide corrected photometric measurements has been implemented in \texttt{Phosphoros}.

A more advanced implementation would involve the creation of posteriors for the $C^\alpha_{T}(\Delta\lambda)$ correction factors. This would have the advantage {of relying} less on a specific best-fit solution, and is probably more robust. However, this makes the output more cumbersome to use, since the flux and uncertainty are replaced by a full posterior distribution. A convenient way to deal with these distributions, including the correlations between the correction factors for different passbands, is to provide a sampling of the $C^\alpha_{T}(\Delta\lambda)$ posteriors.

We note finally that relying on TF to determine the $C^\alpha_{T}(\Delta\lambda)$ does not lead to degeneracies. Indeed, even if two comparable solutions at very different redshifts exist, they would have by definition similar spectral shapes over the range of wavelengths covered by the passbands.  The fact that the templates do not match exactly the SEDs of real objects is not a serious problem either, especially if the full posteriors of the correction factors are used, because what is most relevant is the range of colours provided by the templates.

\subsection{Passband variations in ML algorithms}
{It is not straightforward to apply passband variation corrections to ML algorithms. The main advantage of these approaches is indeed that they do not rely on the knowledge of any SED, which is necessary to compute the photometric corrections. It is possible to determine the SEDs of the objects in the training set with TF, making use of the fact that their true redshifts are known. Passband variations can be taken into account using a procedure where the SED and the photometric corrections are determined iteratively until the procedure converges on consistent SED and corrections.}

{Once the SEDs of the training set objects are obtained, it is possible to correct their photometric measurements to match the colour space of any source in the target sample. However, as anticipated in Sect.~\ref{sec:ml}, most ML algorithms would require a potentially computationally expensive training phase to match every single colour space of the target sample. One notable exception is the $k$-nearest neighbours algorithm \citep[$k$-NN;][]{Cover1967}, which does not require any training. The algorithm finds the $k$ closest reference objects to a given source, and compute the value of interest using some (weighted or not) average over theirs. NNPZ \citep{Tanaka2018} is an example of an implementation of the $k$-NN algorithm for the determination of \phz s. In such a case, it is straightforward apply the $k$-NN by correcting first, for each target, the fluxes of the training set objects based on the correction factors of the target.}

{The $k$-NN algorithm relies on a definition of the distance between two objects. One possibility is to adopt a $\chi^2$ distance, with the likelihood given by $\exp{\left(-\chi^2_\alpha/2\right)}$, where $\chi^2_\alpha$ is given by Eq.~(\ref{eq:chi2}), with the exception that $\alpha$ runs over the training set, instead of a grid of models. In such a case, the correction procedure and all equations from Sect.~\ref{sec:corr} can be used with minimal and straightforward modifications.}

\section{Conclusion}
We studied the effects of variations of photometric passbands in the process of \phdz\ determination. {Biases due to passband variations are expected to be positive due to the usual shift of transmissions towards shorter wavelengths for off-axis sources}. We found that, when taking into account the SED of the objects, it is sufficient to consider only the change in the {mean} wavelengths of the passbands in order to get a very accurate determination of the correction to be applied to the observed flux. Crucially, the mean wavelength can be determined irrespectively of the SED of the source. This simplification allowed us to propose an accurate and efficient correction that can be readily implemented in any TF algorithm, and has already been done in \texttt{Phosphoros}.

While the method we proposed here is able to remove effectively some bias in the \phdz\ predictions, a significant bias remains. In the application to the CFHTLS photometric catalogue, we found a reduction by a factor 2 at $z=0.5$, although it remained a factor 6 higher than the \Euclid requirements. Remaining {sources} of bias were discussed. Bias may originate from the photometry itself. It could also be due to the inadequacy of the model; for instance, the templates might not represent accurately the SEDs of galaxies, or they may lack diversity. But {the bias} could also be due to the application of wrong priors, in particular hidden priors \citep{Schmidt2020}. These sources of bias could in principle be alleviated with deeper understanding of the properties of galaxies, but this remains a very challenging task. {ML algorithms} are immune to some photometric issues and to model-dependent effects, but are particularly sensitive to the hidden priors resulting from the construction of the training sample.

TF {has} different strategies to cope with biases inherent to the photometry itself, {such as} corrections for offset in the zero-points \citep{Ilbert2006}, or template adaptations \citep{Coupon2009}. Each of these methods should be used to remove the bias as much as possible. In many situations, {ML algorithms should be largely insensitive to such biases}. Passband variations, however, induce biases in the photometry that depend on the object, and taking them into account in ML algorithms would require an extremely complex training sample. {Passband variations, if not corrected, can also imprint spurious spatial patterns that can {bias} cosmological parameter estimation; this is especially a concern for photometric galaxy clustering.}

The method proposed here is able to remove {most of the bias} due to time and spatial variations of the passbands. While this is only one of the many biases affecting \phdz\ determination, it is important to perform these corrections, because the cosmological requirements on the accuracy of \phz s are so stringent for a mission such as \Euclid that a post-processing calibration step is required; starting this step with the smallest possible bias is necessary if we want this calibration step to succeed. As a bonus, our method is able to provide corrected photometry, so that it can be used for other applications, for instance, the computation of new \phdz\ predictions based on any other algorithm, in particular using ML.

\begin{acknowledgements}
	\AckEC\\[1mm]
	
	Based on observations obtained with MegaPrime/MegaCam, a joint project of CFHT and CEA/IRFU, at the Canada-France-Hawaii Telescope (CFHT) which is operated by the National Research Council (NRC) of Canada, the Institut National des Science de l'Univers of the Centre National de la Recherche Scientifique (CNRS) of France, and the University of Hawaii. This work is based in part on data products produced at Terapix available at the Canadian Astronomy Data Centre as part of the Canada-France-Hawaii Telescope Legacy Survey, a collaborative project of NRC and CNRS. 
\end{acknowledgements}

\bibliographystyle{aa} 
\bibliography{MyLibrary}

\begin{appendix}

\section{Computation of the correction functions}
\label{sec:corr_factor}
We describe here how the correction factors $C^\alpha_T(\Delta\lambda)$ can be computed for the transmission curve $T(\lambda)$, in order to correct the fluxes in Eq.~(\ref{eq:chi2_corr}), with $\alpha$ being an index that runs through all the {model parameters}. In the case of real reference objects with {known} SEDs, $\alpha$ is just a integer sequence. In the case of TF, $\alpha$ runs through all points in the computed grid, that is, $\alpha\in \{z_i, \mathit{SED}_j, {E_{B-V}^\mathrm{int}}_k\}$, where $i$, $j$ an $k$ are integer sequences that enumerate all redshifts, all SEDs, and all internal reddening values, respectively, in the grid.

\begin{figure}
\includegraphics[width=8.65cm]{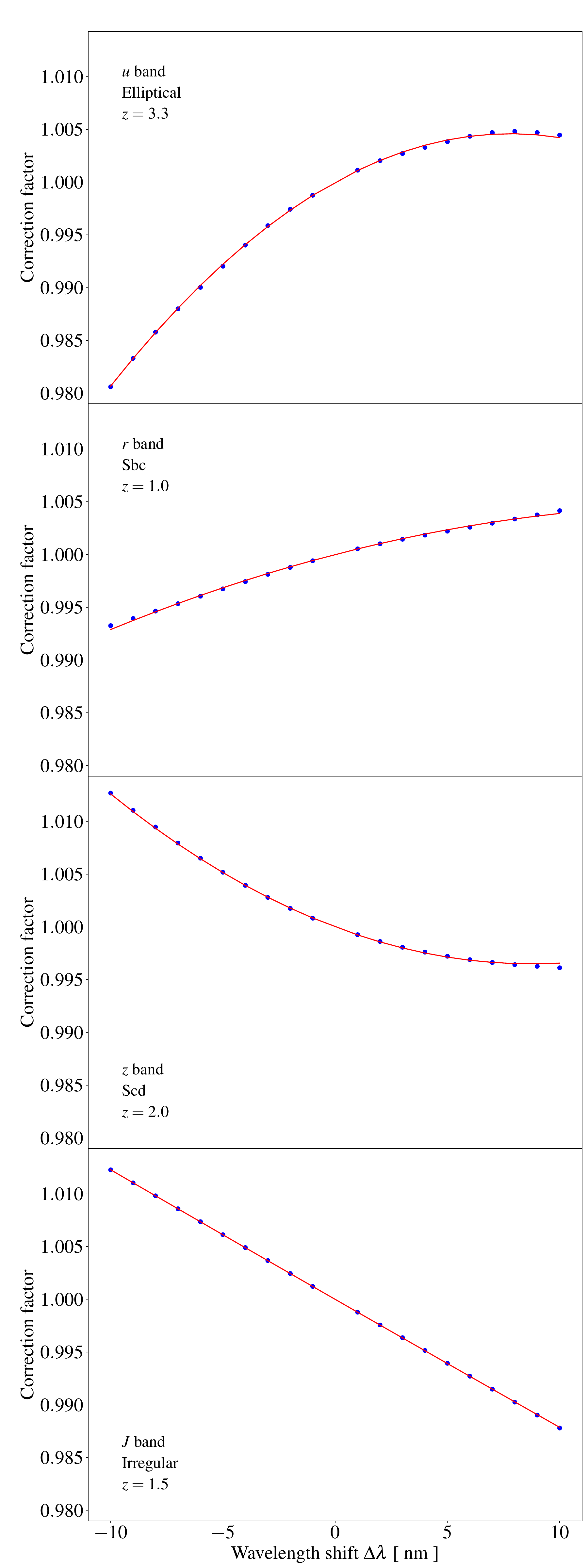}
\caption{\label{fig:corr_factor}Correction factors $C^{\alpha}_T(\Delta\lambda)$ (blue points) and the corresponding approximation with a second-degree polynomial (red line) for four arbitrary values of the passband $T(\lambda)$ and of the coordinates $\alpha$ in the model grid.}
\end{figure}

For each passband and each $\alpha$, we compute the fluxes $F^\alpha_{T}(\Delta\lambda)$ over a sequence of $\Delta\lambda$ that encompasses (but should not largely exceed) the range of shifts in the mean wavelengths for the passband of interest. We use here 20 values between $-10$ and $+10$\,nm. The correction is therefore:
\begin{equation}
C^{\alpha}_T(\Delta\lambda)=F^\alpha_{T}(\Delta\lambda)/F^\alpha_{T}(0)\;.
\end{equation}
We then approximate $C^\alpha_{T}(\Delta\lambda)$ with a polynomial expression. In general, we would like to minimise the number of parameters used to represent $C^\alpha_{T}$, {because the number of models $\alpha$ might be very large, and because high-degree polynomials might cause unwanted fluctuations}. Figure~\ref{fig:corr_factor} shows a few examples of correction factors as a function of $\Delta\lambda$ for different values of $T$ and $\alpha$. Investigating the $C^{\alpha}_T(\Delta\lambda)$ functions, {we find that a second-degree polynomial always provides a very good approximation}, requiring three parameters for each $(T,\alpha)$. However, we have the constraint $C^{\alpha}_T(0)=1$, so that there are effectively only two free parameters. Thus we define
\begin{equation}
\label{eq:chat}
\hat{C}^\alpha_{T}(\Delta\lambda)=\frac{C^{\alpha}_T(\Delta\lambda)-1}{\Delta\lambda}\;.
\end{equation}
We point out that $\hat{C}^\alpha_{T}(\Delta\lambda)$ is well defined for $\Delta\lambda\neq 0$. We can then approximate $\hat{C}^\alpha_{T}(\Delta\lambda)$ with a linear function:
\begin{equation}
\hat{C}^\alpha_{T}(\Delta\lambda)\approx a^\alpha_{T}\Delta\lambda+b^\alpha_{T}
\end{equation}
The coefficients $a^\alpha_{T}$ and $b^\alpha_{T}$ can be obtained using a simple least-square minimisation. Finally, we have, using Eq.~\eqref{eq:chat}:
\begin{equation}
\label{eq:deg2fit}
C^{\alpha}_T(\Delta\lambda)\approx a^{\alpha}_T\Delta\lambda^2+b^{\alpha}_T\Delta\lambda+1\;,
\end{equation}
The second-order approximation from Eq.~(\ref{eq:deg2fit}) is shown on Fig.~\ref{fig:corr_factor} for four arbitrary values of the passband $T$ and of the model parameter $\alpha$. Figure~\ref{fig:sim-bias} shows that using the approximated correction Eq.~(\ref{eq:deg2fit}) is practically {indistinguishable} from computing the flux using the full shifted passband.

\section{Other effects leading to passband variations}
\label{sec:other}
Several other effects may affect the passband, and induce passband variations. However, depending on {the} details of the effects, they may be treated differently. We discuss three such effects below.

\subsection{Galactic reddening}
\label{sec:ebv}
Dust in our Galaxy scatters the light emitted by extragalactic sources. The induced, extrinsic, so-called Galactic reddening depends on the direction in the sky, and its amplitude, parameterised by the value of the reddening $E_{B-V}$, can be determined either from the observation of stars or extragalactic sources with well known intrinsic SEDs, such as quasars or passive galaxies \citep[e.g.,][]{Schlegel1998,Schlafly2010,Schlafly2011,Mortsell2013}, or from direct measurement of {the  quantity of dust}, for instance by combining {\it Planck} data with IRAS $100\,\mu$m data \citep{PlanckCollaboration2014}. Galactic reddening is also strongly wavelength-dependent (hence the name), {and} affects the colours of the observed objects, {so that its wavelength dependence can be determined} from observations of well-characterised objects \citep{Fitzpatrick1999}.

Photometry is often corrected for Galactic reddening by computing a colour term that depends only on the amount of dust in the line of sight. However, the (unknown) SED of the source needs to be taken into account in order to compute an accurate correction. Galactic reddening could be applied to all models in the grid, in order to derive, for all passbands, the shifts in the {mean} wavelength that are induced considering the model and the reddening $E_{B-V}$; however, when computing these shifts, the flux corrections are obtained at the same time. Therefore, in \citet{Galametz2017} a more direct approach has been adopted, where the flux corrections are derived directly, without computing the wavelength shifts. Thus, while Galactic reddening could be corrected by computing the wavelength shifts following the approaches we used here, in \texttt{Phosphoros} the two corrections are computed separately. We point out again that the work of \citet{Galametz2017} is very similar to the approach described here, and has been the inspiration for the wavelength-shift correction presented in this paper.

\subsection{Absorption in the intergalactic medium}
The intergalactic medium (IGM) is filled with hydrogen that very efficiently absorbs the ultraviolet emission of objects at cosmological distances either through Lyman bound-bound transitions, or through the Lyman bound-free transition below 912\,\AA. Because of the clumpy distribution of matter in the Universe, each line of sight is subject to a different amount of absorption. Contrarily to the case of Galactic reddening, the amount of gas in the line of sight cannot be measured, and the SED cannot be recovered exactly. However, an average IGM transmission curve $G_{z}(\lambda_0)$, which depends strongly on the redshift $z$ and on the rest-frame wavelength $\lambda_0=\lambda/(1+z)$, can be determined based on an analytic modelling of the distribution of gas clumps \citep{Madau1995,Inoue2014}, or on cosmological numerical simulations \citep{Meiksin2006}.

The IGM transmission curve $G_{z}(\lambda_0)$ can be seen as modifying the transmission curve $T(\lambda)$, such that the object is observed through an effective curve $T'(\lambda)=G_{z}(\lambda/(1+z))\,T(\lambda)$. However, for TF algorithms, the IGM transmission depends only on the parameters $\alpha$ of the model, so that it can be applied directly to the model SED when computing the predicted fluxes. Therefore, there is no need express the IGM absorption as an effect of passband variation.  

\subsection{Effect of the motion of the Earth}
Light from extragalactic sources is affected by Doppler effects due to the velocity $v$ of the observer. Several components contribute to $v$: the motion of the observatory due to the Earth's rotation, or, in the case of \Euclid, due to the motion of the spacecraft around the second Lagrangian point $L_2$; the revolution of the Earth, or of $L_2$, around the Solar system barycentre; and finally the peculiar motion of the Solar system barycentre with respect to the Universe's comoving frame. Velocities due to Earth's rotation and revolution can be very easily calculated \citep[e.g., using the \texttt{barycorrpy} Python package\footnote{\url{https://pypi.org/project/barycorrpy/}};][]{Kanodia2018}. Earth's rotation induces maximum velocities in the direction of the Earth's equatorial plane below 1\,\kms, that is, $v/c\sim 10^{-6}$, which have negligible effect. Likewise, the motion of \Euclid around $L_2$ has velocities below 1\,\kms. Earth's revolution velocity is however of the order of $30$\,\kms, that is, $v/c\sim 10^{-4}$, reaching a maximum along the ecliptic plane. Finally, the peculiar motion of the Solar system barycentre can be obtained from the dipole of the cosmic microwave background (CMB), and implies a peculiar velocity of $371\pm 1$\,\kms\ towards Galactic coordinates $(\ell,b)=(\ang{264.14;;}\pm0.15,\ang{48.26;;}\pm0.15)$, or $v/c\sim 10^{-3}$ \citep{Fixsen1996}; this direction is also in the ecliptic plane. Velocities must be added as vectors, and only the resulting radial velocity matters (at these velocities, transverse Doppler effect can be ignored). Even though the \Euclid survey will avoid the ecliptic plane because of the zodiacal light \citep{Scaramella2022}, the avoidance angle of \ang{10;;} means that velocities up to 98\% of the maximal velocities of the Earth's revolution and Solar-system peculiar motion will occur.

The resulting radial velocities affect the photometric fluxes through Doppler boosting, which is the combination of three effects: relativistic aberration, time dilation, and blue- or redshifting of the spectrum. To first order, the first two effects lead to an achromatic amplification $(1-3v/c)$, where a positive velocity indicates a motion away from the source. The amplification or attenuation can reach a level of a few $10^{-3}$, or a few millimagnitudes, but it is removed by the calibration process, as it affects all sources in an exposure in the same way.

The third effect results in a shift of the observed wavelengths by a term $(1+v/c)$ (with the same definition of the sign of $v$ as above). This induces approximately a translation of the passband of $(1+v/c)$ times the {mean} wavelength of the passband. Considering the {mean} wavelength of the MegaCam $r$ filter is 630\,nm, the maximum shift due to these velocities is about 0.8\,nm, which represents about 10\% of the  wavelength shift due to the off-axis dependence of the filter, and is therefore non-negligible. Since the radial velocity can be computed easily, the resulting wavelength shift can be applied to the \phdz\ determinations using the algorithm presented here.

\end{appendix}

\end{document}